\documentclass[aps,showpacs,showkeys,floatfix,amsmath,amssymb]{revtex4}
\usepackage[T1]{fontenc}
\usepackage[latin1]{inputenc}
\usepackage{graphicx}
\usepackage[dvips]{color}
\usepackage{colordvi}

\def\co{\hat{c}}
\def\bo{\hat{b}}

\begin{document}

%A SYMMETRY-PROJECTED VARIATIONAL APPROACH TO THE 1-DIMENSIONAL HUBBARD-MODEL

\title{A symmetry-projected variational approach to the 1-dimensional
Hubbard-model}
\author{K.W. Schmid, T. Dahm, J. Margueron and H. M\"uther}
\affiliation{Institut f\"ur Theoretische Physik der Universit\"at T\"ubingen,
Auf der Morgenstelle 14, D-72076 T\"ubingen, Germany}

%\date{submitted in August 2004}

\begin{abstract}
We apply a variational method devised for the nuclear many--body problem
to the 1-dimensional Hubbard--model with nearest neighbor hopping and
periodic boundary conditions. The test wave function consist for each
state out of a single Hartree--Fock determinant mixing all the sites
(or momenta) as well as the spin--projections of the electrons. Total spin
and linear momentum are restored by projection methods before the variation.
It is demonstrated that this approach reproduces the results of exact
diagonalisations for half--filled $N=12$ and $N=14$ lattices not only for the
energies and occupation numbers of the ground but also of the lowest excited
states rather well. Furthermore, a system of 10 electrons in a $N=12$ lattice
is investigated and, finally, a $N=30$ lattice is studied. In addition to
energies and occupation numbers we present the spectral functions computed
with the help of the symmetry--projected wave functions, too. Also here nice
agreement with the exact results (where available) is reached.
\end{abstract}



\maketitle

\section{INTRODUCTION\label{sec1}}

Consider a finite number of $N$ identical Fermions in a model space
defined by some finite number $M$ of suitably chosen orthonormal single
Fermion basis states. Assume furthermore, that the effective Hamiltonian of
these Fermions appropriate for this model space is known. Then, at
least in principle, this many--Fermion problem is exactly solvable.
One only has to distribute the $N$ Fermions over the $M$ orbitals
according to the Pauli principle, i.e., construct all possible Slater--determinants,
and then to diagonalize the known Hamiltonian in the resulting
configuration space of dimension ${M\choose N}$.

\noindent
The nuclear shell--model (see~\cite{bro01} for a recent review and
references therein) is one realization of this scheme. In order to describe
ground-state and low energy excited states of a nucleus  $Z_v$ valence
protons and $N_v$ valence neutrons are distributed over $M_p$ and $M_n$
single-particle basis states and a suitably chosen effective Hamiltonian
is then diagonalized in the resulting configuration space. The dimension of
this configuration space which is
${{M_p}\choose{Z_v}}\cdot{{M_n}\choose{N_v}}$, can, 
however, be reduced considerably, if use is made of the symmetries of the
effective Hamiltonian. So, e.g., the effective nuclear many--body Hamiltonian
is a scalar in normal space and thus does neither mix states with  different
total angular momentum nor with different z--projections of the latter.
Furthermore, neglecting weak interactions, parity becomes a good quantum
number, too. Consequently, the above Slater--determinants can always be coupled
to configurations with definite parity and angular momentum quantum numbers and
the Hamiltonian can then be diagonalized in the much smaller configuration
spaces corresponding to such a symmetry representation. However, even if use
is made of all the symmetries, in general the dimensions are far too large to
be numerically tractable. Thus in most cases one has to rely on approximate
methods, which truncate the complete shell--model expansion of the wave
functions to a numerically feasible number of configurations. How this can be
done without loosing the essential degrees of freedom being relevant for any
particular state under consideration is the central question of nuclear
structure physics.

\noindent
The Hubbard--model, a schematic model developed to describe some basic
features of solid state physics, is another example for the scheme outlined
above. In this case $N_e$ electrons
are distributed over the $N$ sites of a 
lattice and the corresponding model--Hamiltonian is then diagonalized
within the resulting configuration space (see, e.g.~\cite{geb97} and references
therein). Again, the dimension of this configuration space increases
drastically with the number of lattice points $N$ and again, even if all the
available symmetries are used, the dimensions are in general far
too large to allow for an exact solution. Thus again approximate
methods are asked for and rather successful ones have been developed
within the past~\cite{geb97}. The aim of the present study is to adopt an
approximation scheme, which has successfully been applied in nuclear physics,
and apply it to the Hubbard model to test its efficiency for this system.  

\noindent
One of the most successful truncation approaches to the nuclear
many--body problem has been addressed in a recent review article~\cite{sch04}.
It works with general Hartree--Fock--Bogoliubov configurations, breaking
all the symmetries required by the effective nuclear many--body Hamiltonian,
as basic building blocks. These symmetries are restored with the help
of projection techniques and the underlying HFB--transformations as well as
the configuration--mixing are then determined by variational
calculations. In quantum chemistry such kind of approaches are 
called {\it{unrestricted self-consistent field approximations with
variation after projection}} and it is known that electronic correlations
can be accounted for this way by breaking symmetries \cite{Ful95}.

\noindent
Here we want to address the question how well the symmetry projected
variational approach will perform when applied to the Hubbard model. For
simplicity we shall restrict ourselves to the 1--dimensional Hubbard model and
use only general Hartree--Fock instead of Hartree--Fock--Bogoliubov
configurations in the  symmetry projected variational approach. As a test
we will determine energies and single-particle spectral functions and compare
the results to exact results for cases, for which the exact diagonalization can
be done. We will also demonstrate the feasibility of the symmetry projected
variational approach for cases, which cannot be solved by brute force
diagonalization.

\noindent
We present the relevant formalism in sec.~\ref{sec2}. We start with a short
sketch of the 1--dimensional Hubbard model with nearest neighbor hopping and
periodic boundary conditions in sec.~\ref{ssec21}. In sec.~\ref{ssec22} we
shall  then derive the variational equations for the Hartree--Fock approach
with spin-- and linear momentum--projection before the variation for the ground
as well as for the excited states of the considered system. How to calculate
the corresponding (hole-- a well as particle--) spectral functions is discussed
in sec.~\ref{ssec23}. Sec.~\ref{sec3} contains some general properties of the
Hubbard--model and presents then the results obtained within the variational
approach for half--filled $N=12$--, $N=14$-- and $N=30$--lattices. Where
possible these results are compared  with those of complete diagonalisations.
Furthermore, some results for a 10 electron system in a  $N=12$ lattice are
presented. Finally, some conclusions and an outlook are given in
sec.~\ref{sec4}. 

\section{THEORY\label{sec2}}

\subsection{The 1-dimensional Hubbard--model\label{ssec21}}

In its simplest version~\cite{geb97} the effective Hamiltonian of the 
1-dimensional Hubbard--model has the form
\begin{eqnarray}
\hat H\,\equiv\,-t\,\sum\limits_{j=1}^N\,\sum\limits_{\sigma=-1/2}^
{+1/2}\,\Biggl\{\co^{\dagger}_{j+1\sigma}\co_{j\sigma}\,+\,
\co^{\dagger}_{j\sigma}\co_{j+1\sigma}\Biggr\}\,+\,
U\sum\limits_{j=1}^N\,\co^{\dagger}_{j\uparrow}
\co^{\dagger}_{j\downarrow} \co_{j\downarrow} \co_{j\uparrow}\;.\
\label{eq1}
\end{eqnarray}
Here the operator $\co^{\dagger}_{j\sigma}$ creates from the particle
vacuum $\vert 0\rangle$ an electron with spin--projection $\sigma=\pm1/2$
along an arbitrary chosen quantization axis on the site $j$ ($j=1,\dots,\,N$),
the corresponding annihilator $\co_{j\sigma}$ destroys such an electron.
Obviously, these operators fulfill the standard anti--commutation relations
for Fermion--operators
\begin{eqnarray}
\left\{\co^{\dagger}_{j\sigma},\,\co_{j^{\prime}\,\sigma^{\prime}}\right\}
\,\equiv\,\co^{\dagger}_{j\sigma}\,\co_{j^{\prime}\,\sigma^{\prime}}\,+\,
\co_{j^{\prime}\,\sigma^{\prime}}\,\co^{\dagger}_{j\sigma}
\,=\,\delta_{j,\,j^{\prime}}\,\delta_{\sigma,\,\sigma^{\prime}}\;.
\label{eq2}
\end{eqnarray}
The Hamiltonian~(\ref{eq1}) simulates a system of electrons in a periodic potential,
in which each of the $N$ potential wells is supposed to have only a single
bound electron state, which can be occupied by at most two electrons with
opposite spin--directions. In case that two electrons are occupying the same
state, they feel the repulsive Coulomb--interaction ($U>0$). Furthermore, each
electron may tunnel to the neighboring well (if the corresponding site is not
already filled by an electron with the same spin--projection). This
so--called ``nearest--neighbor--hopping'' is described by the
``hopping parameter'' $t>0$. As usual, all energies are measured in units of
this parameter and $N$ is supposed to be an even integer. One furthermore
assumes periodic boundary conditions, i.e. the sites $N+1$ and $1$ are
identical. Thus the system lives on a circle of length $L=N\Delta$. The
spacing $\Delta$ is set to unity in the following. Let us for the moment
consider a ``half--filled'' grid, i.e. $N_e=N$ electrons on
the $N$ sites. It is obvious, that for interaction strength $U=0$ these
electrons form a noninteracting ``Fermi--gas'' in a finite, 1--dimensional box,
while for very large interaction strength $(U\,\to\,\infty)$ an
anti-ferromagnetic ground state with total spin $S=0$ is to be expected~\cite{geb97}.

\noindent
We apply now the Fourier transformation
\begin{eqnarray}
\co^{\dagger}_{\alpha\sigma}\,=\,\frac{1}{\sqrt{N}}\sum\limits_{j=1}^N
\exp\{-ik_{\alpha}\,j\}\,\co^{\dagger}_{j\sigma}
\label{eq3}
\end{eqnarray}
on the basis orbits ($\Delta=1$). This yields a set of $N$ single electron
states in momentum space with momenta
\begin{eqnarray}
k_{\alpha}\,\equiv\,\frac{2\pi}{N}\,\alpha\qquad
\alpha\,=\,-N/2+1,\dots,N/2\,,
\label{eq4}
\end{eqnarray}
in terms of which the Hamiltonian~(\ref{eq1}) gets the form
\begin{eqnarray}
\hat H\,=\,-2t\,\sum\limits_{\alpha}
\sum\limits_{\sigma=-1/2}^{1/2}\,\cos\left(\frac{2\pi}{N}
\,\alpha\right)\,\co^{\dagger}_{\alpha\sigma} \co_{\alpha\sigma}\,+\,\frac{U}{N}
\sum\limits_{\alpha,\,\beta,\,\gamma,\,\delta}\,
\delta^{0,\,\pm N}_{\alpha+\beta-\gamma-\delta}
\co^{\dagger}_{\alpha\uparrow} \co^{\dagger}_{\beta\downarrow}
\co_{\delta\downarrow} \co_{\gamma\uparrow}\,,
\label{eq5}
\end{eqnarray}
where the generalized Kronecker--symbol is one, if
$\alpha+\beta-\gamma-\delta$ is either 0 or $\pm N$ (because of
  the periodic boundary conditions), and zero else. The linear
momentum quantum numbers $\alpha,\,\beta,\,\gamma,\,\delta$
run all over $-N/2+1,\dots\,N/2$. Because of the ``cosine--dispersion'' 
in the one--body term, the single particle spectrum contains one state
with energy $-2t$ (for $\alpha=0$), one state with energy $+2t$ (for
$\alpha=N/2$) and $N-2$ two--fold degenerate states with energies
$\cos([2\pi/N]\alpha)$ (for $\alpha=\pm 1,\dots,\pm(N/2-1)$).

\noindent
Obviously, the Hamiltonian~(\ref{eq1}) or (\ref{eq5}) conserves the total number of 
electrons. Furthermore, it is easy to show, that it commutes with the
square of the total spin operator ${\hat S\,}^2$ as well as with its
3--component ${\hat S}_z$ and is thus a scalar in spin--space. Hence its 
eigenstates can be classified according to the corresponding spin--quantum
number $S$ and are degenerate for all the $2S+1$ values of its z--projection
$\Sigma=-S,...,S$. For an even number of electrons only integer values for the
total spin can occur. Since each of these spin--values has a $\Sigma=0$
component, it is sufficient to diagonalize~(\ref{eq1}) or (\ref{eq5}) in the space of the
$\Sigma=0$ configurations in order to obtain all the eigenstates. 
In contrast, for an odd number of electrons $S$ is a half--integer, so
that in this case all the configurations with $\Sigma=1/2$ have
to be used as basis states. 
The lowest eigenvalues and eigenvectors for such a matrix with
large dimension can be obtained by applying efficient algorithms like e.g. the 
Lanczos--method~\cite{lan}. For half--filling ($N_e=N$ even) the total number of
$\Sigma=0$ configurations to be treated is
\begin{eqnarray}
n(N/2,\,\Sigma=0)\,\equiv\,{N \choose N/2}^2\,,
\label{eq6}
\end{eqnarray}
and thus increases drastically with the number $N$ of available
sites (for $N=16$,e.g., eq.~(\ref{eq6}) yields already 165 636 900 configurations).
Hence the applicability of this approach is rather limited.
\smallskip\noindent
The situation becomes slightly better, if instead of all $\Sigma=0$
Slater--determinants only those configurations with a definite total
spin--value $S$ are included. Again for half--filling, one obtains here
dimensions of
\begin{eqnarray}
n(N/2,\,S)\,\equiv\,{N \choose {N/2 - S}}^2\,-\,{N \choose {N/2 - S -1}}^2\,,
\label{eq7}
\end{eqnarray}
which for $N=16$ still amount to 34 763 300 $S=0$ and even 66 745 536
$S=1$ configurations, respectively. In addition, such a procedure requires
the coupling of the simple Slater--determinant to configurations with 
good total spin and thus complicates the calculation of the various matrix
elements considerably.

\noindent
There is, however, a further symmetry of the Hamiltonian, which can be used
to reduce the dimension of the configuration spaces. As can be shown easily,
the operator~(\ref{eq5}) commutes (modulo $\pm 2\pi$) with the
Hamiltonian of total linear momentum
\begin{eqnarray}
\hat P\,\equiv\,\sum\limits_{\alpha,\,\sigma}\,\left\{
\frac{2\pi}{N}\alpha\right\}\co^{\dagger}_{\alpha\sigma}
\co_{\alpha\sigma}\,,
\label{eq8}
\end{eqnarray}
and thus the eigenstates can be classified in addition by a momentum
quantum number $\xi$, too. The corresponding total momentum is 
$k_{\xi}\,=\,(2\pi/N)\xi$. The total linear momentum quantum number can
assume all values given in eq.~(\ref{eq4}) or, equivalently, all
integer values ($\xi$) in between 0 and $(N-1)$, where use has been made
of the periodic boundary conditions.  In addition, it is immediately seen,
that the Hamiltonian~(\ref{eq5}) remains unchanged, if the signs of
all the linear momentum quantum numbers are flipped (because of the
periodic boundary conditions $-N/2\,\to\,-N/2+N\,=\,N/2$). Thus the states
with $\xi=1$ are degenerate with those with $\xi=N-1$, those with $\xi=2$
with those with $\xi=N-2$, etc., while the states with $\xi=N/2$
(again, since $-N/2\,\to\,-N/2+N=N/2$) occur only once.

\noindent
For half--filling, the total number of configurations with a given total
spin $S$ and given linear momentum quantum number $\xi$ cannot be given
in a closed form but it can be easily calculated numerically. It turns
out that
\begin{eqnarray}
n(N/2,\,S \xi)\,\simeq\,n(N/2,\,S)/N
\label{eq9}
\end{eqnarray}
is a rather good first guess for all possible $\xi$--values, if $N$ is
sufficiently large. For $N=16$ and $S=0$, e.g., the dimensions for the various
configuration spaces are 2 172 400 (for $\xi=1,\,3,\,5,\,7,\,9,\,11,13,\,15$),
2 173 008 (for $\xi=2,\,6,\,10,\,14$), 2 173 016 (for $\xi=4,\,12$) and
2 173 018 (for $\xi=0,\,8$), respectively.

\noindent
However, even if spin and linear momentum are used as ``good'' quantum numbers,
the dimension of the resulting configuration spaces still increases drastically
with the number of sites. For $N=18$ and $S=0$ eq.~(\ref{eq9}) yields already
$2.5*10^7$, and for $N=30$ and $S=0$ even $9.7*10^{13}$ configurations for
each $\xi$ have to be treated. Thus, at least for half--filling and
$N\geq 18$, an exact solution is almost impossible even on modern computers.
This statement remains valid even if we furthermore consider the symmetry 
of the Hubbard Hamiltonian under charge rotation and the corresponding conserved
quantum numbers.

\noindent
Therefore one has to rely on approximate methods, which truncate the
complete expansion of the wave functions to a numerically feasible
number of configurations. How this can be done without loosing the
essential degrees of freedom relevant for the particular states under
consideration is the central question not only for the Hubbard--model 
but for any other finite many--body problem as well.

\noindent
One of the most favored methods for such a truncation is provided
by Greens--function Monte--Carlo calculations, in which the relevant
configurations are selected according to their statistical weight.
This method has been applied rather successfully not only to the
Hubbard--model~\cite{gfmc1} but also to the nuclear many--body 
problem \cite{gfmc2}. It
has, however, the limitation that only expectation values within the
ground state or in some statistical assembly can be obtained. For
Fermion--systems the method suffers in addition from the well--known
``sign--problem". In nuclear physics both these shortcomings could be
overcome in recent years by the so--called 
``Quantum--Monte--Carlo--Diagonalization'' (QMCD) method~\cite{ots01}, in which
again the relevant configurations are selected by stochastic methods,
however, the sign--problem can be bypassed and ground as well as
excited states can be treated on equal footing. 

\noindent
However, stochastic selection is not the only promising method for truncation
in finite many--body problems. Another possibility is provided by variational
methods. Here one uses the most general Slater--determinants (or even
generalized Slater--determinants), which can be constructed within the
considered single particle space, as basic building blocks. Unfortunately,
in general these configurations do  break all the symmetries required by the
chosen many--body Hamiltonian and hence cannot be considered as physical
states but only as an approximation introduced to account for as
much as possible of the correlations in as few as possible configurations.
The required symmetries, however, can be restored with the help of projection
techniques, and the resulting symmetry--projected configurations can then be
used as test wave functions in chains of variational calculations in order
to determine the underlying single particle transformations as well as
the configurations mixing. Such symmetry--projected variational
calculations on the basis of general Hartree--Fock--Bogoliubov configurations
have been applied very successfully to the nuclear many--body problem
within the last two decades (see~\cite{sch04} for a recent review and references
therein), and it could be proven that they work equally well as alternative
approaches like, e.g., the QMCD method.

\noindent
In the present work we want to demonstrate that these symmetry--projected
variational methods are not only useful for the nuclear many--body problem
but can also be applied to other finite many--Fermion problems like
the 1--dimensional Hubbard--model described above.

\subsection{Symmetry projected Hartree--Fock\label{ssec22}}

For this purpose we start by introducing ``quasi--particles'' of the
type
\begin{eqnarray}
\bo^{\dagger}_a\,\equiv\,\sum\limits_{\alpha=-N/2+1}^{N/2}\,
\sum\limits_{\sigma=-1/2}^{+1/2}
D^*_{\alpha\sigma,\,a}\,\co^{\dagger}_{\alpha\sigma}\,,
\label{eq10}
\end{eqnarray}
where $D$ is a linear $(2N\times 2N)$ transformation, which has to be
unitary
\begin{eqnarray} 
D^{\dagger}D\,=\,D D^{\dagger}\,=\,{\bf 1}_{2N}\,,
\label{eq11}
\end{eqnarray}
in order to conserve the Fermion anticommutation relations for the
quasi--particle creators~(\ref{eq10}) and the corresponding
annihilators. Eqs.~(\ref{eq10})
and (\ref{eq11}) describe a general Hartree--Fock (HF) type transformation. It
should be stressed here, that eq.~(\ref{eq10}) could still be generalized by including
linear combinations of the annihilators on the right hand side. This would
result in a so--called Hartree--Fock--Bogoliubov (HFB) transformation
as used, e.g., in the approaches reviewed in ref.~\cite{sch04}. In the present
paper, however, we shall restrict ourselves to the simpler HF--transformations
only. Note, that nevertheless the transformation~(\ref{eq10}) mixes all the linear
momentum quantum numbers as well as the spin--projections of the basis states
(\ref{eq3}).

\noindent
In the usual Hartree--Fock approach one searches then for the optimal
one--determinant representation of the $N_e$--electron ground state
\begin{eqnarray}
\vert D\rangle\;=\;\left\{\prod_{h=1}^{N_e}\;\bo^{\dag}_h\;\right\}
\vert 0\rangle\,,
\label{eq12}
\end{eqnarray}
in which the energetically lowest $N_e$ states $\bo^{\dag}_h$ ($h=1,\dots,N_e$) of
the form~(\ref{eq10}) are occupied and the remaining $2N-N_e$ states $\bo^{\dag}_p$ 
($p=N_e+1,\dots,2N$) are empty. In the following the notation 
$h,\,h^{\prime},\dots$ is always used for the occupied states, while
the unoccupied ones will be denoted by $p,\,p^{\prime},\dots$, respectively.

\noindent
Obviously, the determinant~(\ref{eq12}) conserves the total number of electrons
($N_e$) but is neither an eigenstate of the square of the total spin
operator ${\hat S\,}^2$ nor of its 3--component ${\hat S}_z$ nor of the
total linear momentum operator~(\ref{eq8}). These symmetries have therefore to be
restored with the help of projection techniques. For the spin--quantum numbers
this can be achieved via Villars'~\cite{vil66} famous projection operator
\begin{eqnarray}
\hat P^S_{\Sigma,\Sigma^{\prime}}\,\equiv\,
\frac{2S+1}{8\pi^2}\int\,d\Omega\,{D^{S^*}_{\Sigma,\Sigma^{\prime}}\,}
(\Omega)\,{\hat R}_S(\Omega)\,,
\label{eq13}
\end{eqnarray}
where ${\hat R}_S(\Omega)$ is the rotation operator in spin--space,
the Wigner--function $D^S_{\Sigma,\Sigma^{\prime}}(\Omega)$ its 
representation in spin--eigenstates and the integral has to be taken over the
full range of the three Euler--angles. Because of the non--abelian nature of
the rotation group, (\ref{eq13}) is not a true projector in the strict mathematical
sense. In order to achieve independence of the choice of direction for the 
``intrinsic'' quantization axis one is forced to use the linear combination
\begin{eqnarray}
\vert D;\,N_e S \Sigma\rangle\,=\,\sum\limits_{\Sigma^{\prime}=-S}^{+S}\,
\hat P^S_{\Sigma,\Sigma^{\prime}}\vert D\rangle\,f_{\Sigma^{\prime}}
\label{eq14}
\end{eqnarray}
as test wave function in the variation with the mixing coefficients $f$
to be treated as additional variational variables.

\noindent
In a similar way the total linear momentum can be restored. This is done
via the operator
\begin{eqnarray}
\hat C(\xi)\,\equiv\,\frac{1}{N}\sum\limits_{j=1}^N\,
\exp\left\{i\left(\hat P - \frac{2\pi}{N}\xi\right) j\right\}\,,
\label{eq15}
\end{eqnarray}
which projects the determinant (\ref{eq12}) on the component
with linear momentum $k=(2\pi/N)\xi$. 
The operator~(\ref{eq15}) is the finite, 1--dimensional limit of the general operator
restoring Galilean invariance discussed, e.g., in ref.~\cite{sch04}. 
In contrast to nuclear systems where the Galilean principle of
relativity imposes $k=0$, lattice systems allow solutions with
$k>0$: the Hamiltonian~(\ref{eq1}) or (\ref{eq5}) has to be
considered on the 
(infinitely heavy) background of the ions providing the periodic potential.
This background can absorb any change of linear momentum of the electrons
easily so that Galilean invariance for the total system is always ensured. It
should be stressed furthermore, that the spin--projection and the linear
momentum projection have to be performed {\it before} the variation.
Then the correct moment of inertia and mass is restored~\cite{rin80}.

\noindent
Using (\ref{eq15}) in addition to (\ref{eq14}) we obtain the
  projected determinant
\begin{eqnarray}
\vert D;\,N_e \xi S \Sigma\rangle\,=\,\sum\limits_{\Sigma^{\prime}=
-S}^{+S}\,\hat P^S_{\Sigma,\Sigma^{\prime}}\hat C(\xi)\vert D\rangle\,
f_{\Sigma^{\prime}}
\label{eq16}
\end{eqnarray}
as ansatz for our test wave function. The corresponding energy functional
\begin{eqnarray}
E\,\equiv\,\frac{\langle D;\,N_e \xi S (\Sigma)\vert\hat H
\vert D;\,N_e \xi S (\Sigma)\rangle}{\langle D;\,N_e \xi S (\Sigma)\vert
D;\,N_e \xi S (\Sigma)\rangle}\,,
\label{eq17}
\end{eqnarray}
where the spin--projection $\Sigma$ has been put in parentheses since
the total energy does not depend on this quantum number,
has now to be minimized with respect to the mixing coefficients $f$ as well
as to the underlying HF--transformation $D$. 

\noindent
Variation with respect to the $f$'s yields the generalized eigenvalue problem
\begin{eqnarray}
(H\,-\,EN)f\,=\,0\,,
\label{eq18}
\end{eqnarray}
with the constraint
\begin{eqnarray}
f^{\dagger}Nf\,=\,{\bf 1}_{2S+1}
\label{eq19}
\end{eqnarray}
ensuring the orthonormality of the solutions, and the
$(2S+1)\times(2S+1)$--dimensional square matrices $N$ and $H$ given by
\begin{eqnarray}
N_{\Sigma,\,\Sigma^{\prime}}\,\equiv\,
\langle D\vert\hat P^S_{\Sigma,\Sigma^{\prime}}\hat C(\xi)\vert D\rangle
\label{eq20}
\end{eqnarray}
and
\begin{eqnarray}
H_{\Sigma,\,\Sigma^{\prime}}\,\equiv\,
\langle D\vert\hat H
\hat P^S_{\Sigma,\Sigma^{\prime}}\hat C(\xi)\vert D\rangle\,,
\label{eq21}
\end{eqnarray}
respectively. The matrices $N$ and $H$ are hermitian and the
overlap matrix~(\ref{eq20}) is furthermore positive definite. In
eqs.~(\ref{eq20}) and (\ref{eq21}) the obvious quantum 
numbers $S$ and $\xi$ have been suppressed. In the following only the
energetically lowest solution of eq.~(\ref{eq18}) is kept. 

\noindent
The minimization of the energy functional~(\ref{eq17}) with respect to arbitrary
variations of the underlying HF-- transformation $D$ is more involved.
This transformation has to be unitary and thus not all
of the $2N\times 2N$ matrix elements of $D$ are linearly independent.
Nevertheless, an unconstrained minimization of the functional~(\ref{eq17})
can still be performed, if one parameterizes the underlying HF--transformation
via Thouless' theorem~\cite{tho60}, which states that any HF--determinant
$\vert D_d\rangle$ can be represented in terms of the creation and
annihilation operators of some reference
determinant $\vert D_0\rangle$ via
\begin{eqnarray}
\vert D_d\rangle\;=\;c(d)\exp\left\{\sum_{p,h}\; d_{ph}
\bo^{\dag}_p(D_0) \bo_h(D_0)\right\}\vert D_0\rangle\,,
\label{eq22}
\end{eqnarray}
provided that the two determinants are non--orthogonal, since
\begin{eqnarray}
c(d)\;=\;\langle D_0\vert D_d\rangle\,.
\label{eq23}
\end{eqnarray}
The creation operators belonging to the
HF--determinant $\vert D_d\rangle$ are then related to those of the
reference determinant $\vert D_0\rangle$ via
\begin{eqnarray}
\bo^{\dag}_h(D_d)\;=\;\sum_{h'} [L^{-1}]_{hh'}\left(
\bo^{\dag}_{h'}(D_0)\;+\;\sum_{p'}\;d_{p'h'} \bo^{\dag}_{p'}(D_0)
\right)
\label{eq24}
\end{eqnarray}
for the occupied and
\begin{eqnarray}
\bo^{\dag}_p(D_d)\;=\;\sum_{p'} [M^{-1}]_{pp'}\left(
\bo^{\dag}_{p'}(D_0)\;-\;\sum_{h'}\;d^*_{p'h'} \bo^{\dag}_{h'}(D_0)
\right)
\label{eq25}
\end{eqnarray}
for the unoccupied states, respectively. They are now given
in terms of the $(2N-N_e)\cdot N_e$ linear independent variables $d_{ph}$.
The lower triangular $(N_e\times N_e)$ matrix $L$ in (\ref{eq24}) is defined by the
expression
\begin{eqnarray}
{\bf 1}_{N_e}\;+\;d^T d^*\;=\;LL^{\dag}\,,
\label{eq26}
\end{eqnarray}
while the lower triangular $((2N-N_e)\times (2N-N_e))$ matrix $M$ out of (\ref{eq25})
can be obtained from the solution of the equation
\begin{eqnarray}
{\bf 1}_{2N-N_e}\;+\;d^* d^T\;=\;MM^{\dag}\,.
\label{eq27}
\end{eqnarray}
Both, eqs.~(\ref{eq26}) and (\ref{eq27}), are usual Cholesky decompositions.

\noindent
The variational equations resulting from the minimization of the energy
functional~(\ref{eq17}) with respect to the HF--transformation now assume the form
\begin{eqnarray}
\frac{\partial E}{\partial d_{ph}}\;=\;
\left[{M^{-1}}^{\dag}\; G\; L^{-1}\right]_{ph}\;\equiv\;0\,,
\label{eq28}
\end{eqnarray}
where the $((2N-N_e)\times N_e)$ matrix $G$ is defined by
\begin{eqnarray}
G_{ph}\;\equiv\;\sum\limits_{\Sigma,\,\Sigma^{\prime}}\,f^*_{\Sigma}\,
\langle D\vert(\hat H\,-\,E{\bf 1})\hat P^S_{\Sigma,\Sigma^{\prime}}
\hat C(\xi)\,\bo^{\dagger}_p(D)
\,\bo_h(D)\vert D\rangle\,f_{\Sigma^{\prime}}\,.
\label{eq29}
\end{eqnarray}
Once one has reached the solution not only the ``global"
gradient vector~(\ref{eq28}) but 
also the ``local'' one~(\ref{eq29}) does vanish identically. This vanishing of the 
gradient vector~(\ref{eq29}) is a sort of ``generalized Brillouin theorem"~\cite{tho60}~:
it describes the stability of the symmetry--projected  HF--solution with
respect to arbitrary symmetry--projected one particle--one hole excitations.

\noindent
For any given $S$ and $\xi$, the simultaneous solution of the set of
 eqs.~(\ref{eq18}), (\ref{eq19}) and (\ref{eq28}) yields 
 the optimal representation of the energetically
lowest state by one single symmetry--projected HF--type configuration. 
This corresponds to the so--called VAMPIR ({\bf V}ariation {\bf A}fter
{\bf M}ean--field {\bf P}rojection {\bf I}n {\bf R}ealistic model spaces)
approach out of ref.~\cite{sch04}, though restricted here to a HF--type instead of
HFB--type transformations. Since $D$ (and hence $d$) as well as $f$ are
essentially complex matrices this solution results from the minimization of
a function of $m=2\cdot(2N-N_e)\cdot N_e\cdot$ real variables, if even
$N_e$ and $S=0$ is considered, while in general for $S>0$ and arbitrary
$N_e$, $(4S+1)+m$ real variables have to be treated. For this minimization
a quasi--Newton procedure (see, e.g.,~\cite{bro,nr}) is used.

\noindent
Excited states with the same quantum numbers $S$ and $\xi$ can be treated
in a similar fashion, if one ensures the orthogonality with
respect to the solutions already obtained. For the first excited state
($n=2)$ this can be achieved with the help of the projection operator
\begin{eqnarray}
\hat T_{n-1}\,\equiv\,\sum\limits_{i=1}^{n-1}
\vert D^i;\,N_e \xi S \Sigma\rangle
\langle D^i;\,N_e \xi S \Sigma\vert\,,
\label{eq30}
\end{eqnarray}
which projects on the lowest solution ($D^1,\,f^1$) of the form~(\ref{eq16}). For the
variational calculation then instead of (\ref{eq16}) we use its complement
\begin{eqnarray}
\vert D^n;\,N_e \xi S \Sigma\rangle\,\equiv\,(1-\hat T)
\sum\limits_{\Sigma^{\prime}=-S}^{+S}
\,\hat P^S_{\Sigma,\Sigma^{\prime}}\hat C(\xi)\vert D^n\rangle\,
f^n_{\Sigma^{\prime}}
\label{eq31}
\end{eqnarray}
as a test wave function for the variational calculation. The 
procedure~(\ref{eq30}), 
(\ref{eq31}) can be repeated for the second excited state ($n=3$), etc., up to
the $n$ lowest (orthonormal) states. Finally,
the residual interaction between these $n$ states is diagonalized.
This corresponds closely to the EXCITED VAMPIR approach from ref.~\cite{sch04}.

\noindent
Furthermore, we would like to stress, that if a description by one
single symmetry--projected determinant for each state is not sufficient,
further correlations can easily be accounted for by successive
variational calculations for additional determinants as done in the
FED (from {\bf FE}w {\bf D}eterminants) EXCITED VAMPIR approach out of
ref.~\cite{sch04}. This, however, has not been done in the present
work since one determinant gives already a good accuracy, as we
  will see below.

\noindent
Left to be computed are now the symmetry projected
matrices~(\ref{eq20}) 
and (\ref{eq21}) generalized to two different determinants $\vert D^i\rangle$
and $\vert D^k\rangle$ on both sides, because of the eventual use of
(\ref{eq31}) instead of (\ref{eq16}) as test wave function and of more than one determinant
for the description of the $N_e\pm 1$--electron systems as discussed in
sec.~\ref{ssec23}. Furthermore, the corresponding gradient vectors occurring in
eq.~(\ref{eq29}) have to be calculated. One obtains successively
\begin{eqnarray}
\langle D^i\vert\hat P^S_{\Sigma,\Sigma^{\prime}}\hat C(\xi)
\vert D^k\rangle &=&
\frac{1}{N}\sum_{j=1}^N\exp\left\{-i\frac{2\pi}{N}\xi j\right\}\frac{2S+1}
{8\pi^2}\int\,
d\Omega\,{D^{S^*}_{\Sigma,\Sigma^{\prime}}\,}(\Omega)\,
n^{ik}(\Omega,\,j)\,,
\label{eq32} \\
\langle D^i\vert\hat H\hat P^S_{\Sigma,\Sigma^{\prime}}
\hat C(\xi)\vert D^k\rangle &=&
\frac{1}{N}\sum_{j=1}^N\exp\left\{-i\frac{2\pi}{N}\xi j\right\}
\frac{2S+1}{8\pi^2}\int\,
d\Omega\,{D^{S^*}_{\Sigma,\Sigma^{\prime}}\,}(\Omega)\,\cdot\cr
&&\cdot\,h^{ik}(\Omega,\,j)n^{ik}(\Omega,\,j)\,, 
\label{eq33}
\end{eqnarray}
and, 
\begin{eqnarray}
\langle D^i\vert\hat P^S_{\Sigma,\Sigma^{\prime}}\hat C(\xi)\,
\bo^{\dagger}_p(D^k)\,\bo_h(D^k)\vert D^k\rangle &=&
\frac{1}{N}\sum_{j=1}^N\exp\left\{-i\frac{2\pi}{N}\xi j\right\}
\frac{2S+1}{8\pi^2}\int\,
d\Omega\,{D^{S^*}_{\Sigma,\Sigma^{\prime}}\,}(\Omega)\,\cdot\cr
&&\cdot\,n^{ik}_{ph}(\Omega,\,j)n^{ik}(\Omega,\,j)\,,
\label{eq34} \\
\langle D^i\vert\hat H\hat P^S_{\Sigma,\Sigma^{\prime}}
\hat C(\xi)\,\bo^{\dagger}_p(D^k)\,\bo_h(D^k)\vert D^k\rangle\,&=&\,
\frac{1}{N}\sum_{j=1}^N\exp\left\{-i\frac{2\pi}{N}\xi j\right\}
\frac{2S+1}{8\pi^2}\int\,
d\Omega\,{D^{S^*}_{\Sigma,\Sigma^{\prime}}\,}(\Omega)\,\cdot\cr
&&\cdot\,h^{ik}_{ph}(\Omega,\,j)n^{ik}(\Omega,\,j)\,.
\label{eq35}
\end{eqnarray}
In these expressions the spin--rotated and shifted overlap--functions
are given by
\begin{eqnarray}
n^{ik}(\Omega,\,j)\,\equiv\,{\rm det}_{N_e} X^{ik}(\Omega,\,j)\,,
\label{eq36}
\end{eqnarray}
and
\begin{eqnarray}
n^{ik}_{ph}(\Omega,\,j)\,\equiv\,
\sum\limits_{h^{\prime}\in\vert D^i\rangle}
[X^{ik}_{h,\,h^{\prime}}(\Omega,\,j)]^{-1}\,
X^{ik}_{h^{\prime},\,p}(\Omega,\,j)\,,
\label{eq37}
\end{eqnarray}
respectively. The determinant in eq.~(\ref{eq36}) has to be taken for the occupied
$N_e\times N_e$--dimensional part of the ($2N\times 2N$)--matrix $X^{ik}$
\begin{eqnarray}
X^{ik}_{ab}\,\equiv\,\sum\limits_{\alpha,\,\sigma\sigma^{\prime}}
D^i_{\alpha\sigma,\,a}\,S_{\alpha\sigma,\,\alpha\sigma^{\prime}}(\Omega,\,j)\,
D^{k^*}_{\alpha\sigma^{\prime},\,b}
\label{eq38}
\end{eqnarray}
and in eq.~(\ref{eq37}) the indices $h$ and $p$ run over all the
occupied and unoccupied states 
of the HF--determinant $\vert D^k\rangle$, respectively. $(X^{ik})^{-1}$
denotes the inverse of the occupied part of the matrix (38) and
\begin{eqnarray}
S_{\alpha\sigma,\,\alpha\sigma^{\prime}}(\Omega,\,j)\,\equiv\,
D^{1/2}_{\sigma,\,\sigma^{\prime}}(\Omega)\,
\exp\left\{i\frac{2\pi}{N}\alpha\,j\right\}\,.
\label{eq39}
\end{eqnarray}
%For the spin--rotated and shifted matrix elements of the Hamiltonian
%in eq.~(\ref{eq33}) one obtains on the other hand
In eq.~(\ref{eq33}) the spin--rotated and shifted matrix elements
of the Hamiltonian involve also
\begin{eqnarray}
h^{ik}(\Omega,\,j)\,\equiv\,\frac{1}{2}\left\{
t^{ik}(\Omega,\,j)\,+\,\sum\limits_{\alpha\sigma,\,\gamma\sigma^{\prime}}
{\tilde\Gamma}^{ik}_{\alpha\sigma,\,\gamma\sigma^{\prime}}(\Omega,\,j)
{\tilde\rho}^{ki}_{\gamma\sigma^{\prime},\,\alpha\sigma}(\Omega,\,j)
\right\}\,,
\label{eq40}
\end{eqnarray}
where
\begin{eqnarray}
t^{ik}(\Omega,\,j)\,&\equiv&\,\sum\limits_{\alpha\sigma}\left[
-2t\cos\left(\frac{2\pi}{N}\alpha\right)\right]
{\tilde\rho}^{ki}_{\alpha\sigma,\,\alpha\sigma}(\Omega,\,j)\,,
\label{eq41}\\
{\tilde\Gamma}^{ik}_{\alpha\sigma,\,\gamma\sigma^{\prime}}
(\Omega,\,j)\,&\equiv&\,\delta_{\sigma,\,\sigma^{\prime}}\Biggl\{
\delta_{\alpha,\,\gamma}
\left[
-2t\cos\left(\frac{2\pi}{N}\alpha\right)\right]\,+\,
\sum\limits_{\beta\delta}\frac{U}{N}\delta^{0;\pm N}_{\alpha +\beta -\gamma
-\delta}
{\tilde\rho}^{ki}_{\delta -\sigma,\,\beta -\sigma}(\Omega,\,j)\Biggr\}
\nonumber \\
&&-\,(1-\delta_{\sigma,\,\sigma^{\prime}})\Biggl\{
\sum\limits_{\beta\delta}\frac{U}{N}\delta^{0;\pm N}_{\alpha +\beta -\gamma
-\delta}
{\tilde\rho}^{ki}_{\delta\sigma,\,\beta\sigma}(\Omega,\,j)\Biggr\}
\,,
\label{eq42}
\end{eqnarray}
and
\begin{eqnarray}
{\tilde\rho}^{ki}_{\gamma\sigma^{\prime},\,\alpha\sigma}(\Omega,\,j)
\,\equiv\,\sum\limits_{\sigma^{\prime\prime}}
S_{\gamma\sigma^{\prime},\,\gamma\sigma^{\prime\prime}}(\Omega,\,j)
\sum\limits_{hh^{\prime}} D^{k^*}_{\gamma\sigma^{\prime\prime},\,h}
[X^{ik}_{h,\,h^{\prime}}(\Omega,\,j)]^{-1}
D^i_{\alpha\sigma,\,h^{\prime}}\,.
\label{eq43}
\end{eqnarray}
Furthermore, introducing the definitions
\begin{eqnarray}
y^{ki}_{h,\,\alpha\sigma}(\Omega,\,j)\,\equiv\,
\sum\limits_{h^{\prime}} [X^{ik})^{-1}_{h,\,h^{\prime}}(\Omega,\,j)]^{-1}
D^i_{\alpha\sigma,\,h^{\prime}}
\label{eq44}
\end{eqnarray}
for all the occupied states $h$ in the Slater--determinant $\vert D^k\rangle$,
and
\begin{eqnarray}
{\tilde\omega}^{ki}_{\gamma\sigma^{\prime},\,p^{\prime}}(\Omega,\,j)
\,\equiv\,\sum\limits_{\delta\sigma^{\prime\prime}}\,
[{\bf 1}\,-\,{\tilde\rho}^{ki}(\Omega,\,j)]_{\gamma\sigma^{\prime},\,
\delta\sigma^{\prime\prime}}\sum\limits_{\sigma^{\prime\prime\prime}}
S_{\delta\sigma^{\prime\prime},\,\delta\sigma^{\prime\prime\prime}}
(\Omega,\,j) D^{k^*}_{\delta\sigma^{\prime\prime\prime},\,p^{\prime}}
\label{eq45}
\end{eqnarray}
for all the unoccupied states $p^{\prime}$ belonging again to the
transformation $D^k$, one can write the spin--rotated and shifted energy
function out of eq.~(\ref{eq35}) as
\begin{eqnarray}
h^{ik}_{ph}(\Omega,\,j)\,\equiv\,
n^{ik}_{ph}(\Omega,\,j) h^{ik}(\Omega,\,j)\,+\,
\sum\limits_{\alpha\sigma,\,\gamma\sigma^{\prime}}
y^{ki}_{h,\,\alpha\sigma}(\Omega,\,j)
{\tilde\Gamma}^{ik}_{\alpha\sigma,\,\gamma\sigma^{\prime}}(\Omega,\,j)
{\tilde\omega}^{ki}_{\gamma\sigma^{\prime},\,p}
(\Omega,\,j)\,.
\label{eq46}
\end{eqnarray}
The above described variational procedure will be denoted by LMSPHF
(from {\bf L}inear {\bf M}omentum and {\bf S}pin {\bf P}rojected
{\bf H}artree--{\bf F}ock) in the following.

\noindent
Finally, it should be stressed that this procedure can be used for any
number of electrons $N_e=1,\dots,2N$. As we already mentioned, for 
odd $N_e$ the spin quantum numbers $S$ and $\Sigma$ are
half--integer numbers.

\subsection{Spectral functions\label{ssec23}}

Let us now assume that we have obtained the ground state for an even
number of electrons $N_e$ on a lattice with $N$ sites solving the
variational equations out of the sec.~\ref{ssec22}. As we shall see in
sec.~\ref{sec3}, this ground state has always total spin $S=0$ but not
necessarily linear momentum quantum number $\xi_0=0$. 
For instance, in the case of half--filling and $N/2$ even the
  ground state is obtained for $\xi_0=N/2$. 
The underlying HF--transformation will be denoted by $D^1$ and the
total energy by $E_0$.
For $S=0$, there exists only a single coefficient $f_0$, which is
uniquely determined by the normalization
\begin{eqnarray}
f_0\,\equiv\,n_0^{-1/2}\,=\,\langle D^1\vert\hat P^0_{0,0}\hat C(\xi_0)
\vert D^1\rangle^{-1/2}\,.
\label{eq47}
\end{eqnarray}
The projected ground state of the $N_e$--electron system can thus be written as
\begin{eqnarray}
\vert D^1;\,N_e \xi_0 S=\Sigma=0\rangle\,=\,
{\hat P}^0_{0,\,0}\,\hat C(\xi_0)\,\vert D^1\rangle\,
n_0^{-1/2}\,.
\label{eq48}
\end{eqnarray}
We shall now approximate the ($N_e-1$)--electron system 
(characterized by $S=1/2$ and $\xi_{-1}$)
by
\begin{eqnarray}
\vert\tilde h\,;\,N_e-1\,\xi_{-1}\,S=1/2\;\sigma\rangle\,\equiv\,
\sum\limits_{i=1}^m\sum\limits_{h=1}^{N_e}
\sum\limits_{\sigma^{\prime}=-1/2}^{1/2}\,
\hat P^{1/2}_{\sigma,\,\sigma^{\prime}}\hat C(\xi_{-1})
\bo_h(D^i)\vert D^i\rangle\,f_{ih\sigma^{\prime},\,{\tilde h}}\,,
\label{eq49}
\end{eqnarray}
where $\vert D^1\rangle$ refers to the HF--determinant for the
$N_e$--electron ground state solution, while $\vert D^i\rangle$ for
$i=2,\dots,m$ refer to the determinants obtained for the lowest $m-1$ excited
states (which may correspond to different spin-- and linear momentum
quantum numbers than the ground state one's). The mixing
coefficients $f$ are then obtained by solving a generalized
eigenvalue problem similar to eq.~(\ref{eq18})
\begin{eqnarray}
(H\,-\,E_{\tilde h}N)f\,=\,0
\label{eq50}
\end{eqnarray}
with the constraint
\begin{eqnarray}
f^{\dagger}Nf\,=\,{\bf 1}_{2m\cdot N_e}\,.
\label{eq51}
\end{eqnarray}
The overlap-- and Hamiltonian--matrices are given by
\begin{eqnarray}
N_{ih\sigma,\,kh^{\prime}\sigma^{\prime}}\,\equiv\,
\frac{1}{N}&\sum\limits_{j=1}^N \exp\left\{-i\frac{2\pi}{N}\xi_{-1}\,j\right\}
\frac{2}{8\pi^2}\int d\Omega
D^{{1/2}^*}_{\sigma,\,\sigma^{\prime}}(\Omega)\,\cdot\cr
&\cdot\,n^{ik}(\Omega,\,j)n^{ik}_{h,\,h^{\prime}}(\Omega,\,j)\,
\label{eq52}
\end{eqnarray}
and
\begin{eqnarray}
H_{ih\sigma,\,kh^{\prime}\sigma^{\prime}}\,\equiv\,
\frac{1}{N}&\sum\limits_{j=1}^N \exp\left\{-i\frac{2\pi}{N}\xi_{-1}\,j\right\}
\frac{2}{8\pi^2}\int d\Omega
D^{{1/2}^*}_{\sigma,\,\sigma^{\prime}}(\Omega)\,\cdot\cr
&\cdot\,n^{ik}(\Omega,\,j)h^{ik}_{h,\,h^{\prime}}(\Omega,\,j)\,,
\label{eq53}
\end{eqnarray}
where the short--hand notations in the integrands are defined by
\begin{eqnarray}
n^{ik}_{h,\,h^{\prime}}(\Omega,\,j)\,\equiv\,
[X^{ik}_{h^{\prime},\,h}(\Omega,\,j)]^{-1}
\label{eq54}
\end{eqnarray}
and
\begin{eqnarray}
h^{ik}_{h,\,h^{\prime}}(\Omega,\,j)\,\equiv\,
[X^{ik}_{h^{\prime},\,h}(\Omega,\,j)]^{-1}
h^{ik}(\Omega,\,j)\,-\,\sum\limits_{\alpha\sigma,\,\gamma\sigma^{\prime}}
y^{ki}_{h^{\prime},\,\alpha\sigma}(\Omega,\,j)
{\tilde\Gamma}^{ik}_{\alpha\sigma,\,\gamma\sigma^{\prime}}(\Omega,\,j)
z^{ki}_{\gamma\sigma^{\prime},\,h}(\Omega,\,j)\,,
\label{eq55}
\end{eqnarray}
respectively. Here
\begin{eqnarray}
z^{ki}_{\gamma\sigma^{\prime},\,h}(\Omega,\,j)\,\equiv\,
\sum\limits_{h^{\prime\prime},\,\sigma^{\prime\prime}}
S_{\gamma\sigma^{\prime},\,\gamma\sigma^{\prime\prime}}(\Omega,\,j)
D^{k^*}_{\gamma\sigma^{\prime\prime},\,h^{\prime\prime}}
[X^{ik}_{h^{\prime\prime},\,h}(\Omega,\,j)]^{-1}\,,
\label{eq56}
\end{eqnarray}
and all the other functions are defined in sec.~\ref{ssec22}. 
For each possible linear momentum quantum number $\xi_{-1}\;$
the eqs.~(\ref{eq50}) and (\ref{eq51}) yield 
$2m\cdot N_e$ solutions $\tilde h$ with $S=1/2$ and energies $E_{\tilde h}$.

\noindent
Similarly, the $N_e+1$--electron system (characterized by $S=1/2$ and $\xi_{+1}$ ) 
will be approximated by
\begin{eqnarray}
\vert\tilde p\,;\,N_e+1\,\xi_{+1}\,S=1/2\,\sigma\rangle\,\equiv\,
\sum\limits_{i=1}^m\sum\limits_{p=N_e+1}^{2N}
\sum\limits_{\sigma^{\prime}=-1/2}^{1/2}\,
\hat P^{1/2}_{\sigma,\,\sigma^{\prime}}\hat C(\xi_{+1})
\bo^{\dagger}_p(D^i)\vert D^i\rangle\,
g_{ip\sigma^{\prime},\,{\tilde p}}\,,
\label{eq57}
\end{eqnarray}
where again $\vert D^1\rangle$ refers to the HF--determinant for the
$N_e$--electron ground state solution, while the $\vert D^i\rangle$ for 
$i=2,\dots,m$ are taken from the lowest $m-1$ excited states. The mixing
coefficients $g$ are obtained by solving a generalized
eigenvalue problem
\begin{eqnarray}
(H\,-\,E_{\tilde p}N)g\,=\,0
\label{eq58}
\end{eqnarray}
with the constraint
\begin{eqnarray}
g^{\dagger}Ng\,=\,{\bf 1}_{2m\cdot (2N-N_e)}\,.
\label{eq59}
\end{eqnarray}
The overlap-- and Hamiltonian matrices are given by
\begin{eqnarray}
N_{ip\sigma,\,kp^{\prime}\sigma^{\prime}}\,\equiv\,
\frac{1}{N}&\sum\limits_{j=1}^N\exp\left\{-i\frac{2\pi}{N}\xi_{+1}\,j\right\}
\frac{2}{8\pi^2}\int d\Omega
D^{{1/2}^*}_{\sigma,\,\sigma^{\prime}}(\Omega)\,\cdot\cr
&\cdot\,n^{ik}(\Omega,\,j)n^{ik}_{p,\,p^{\prime}}(\Omega,\,j)\,
\label{eq60}
\end{eqnarray}
and
\begin{eqnarray}
H_{ip\sigma,\,kp^{\prime}\sigma^{\prime}}\,\equiv\,
\frac{1}{N}&\sum\limits_{j=1}^N \exp\left\{-i\frac{2\pi}{N}\xi_{+1}\,j\right\}
\frac{2}{8\pi^2}\int d\Omega
D^{{1/2}^*}_{\sigma,\,\sigma^{\prime}}(\Omega)\,\cdot\cr
&\cdot\,n^{ik}(\Omega,\,j)h^{ik}_{p,\,p^{\prime}}(\Omega,\,j)\,,
\label{eq61}
\end{eqnarray}
where the short--hand notations in the integrands are defined by
\begin{eqnarray}
n^{ik}_{p,\,p^{\prime}}(\Omega,\,j)\,\equiv\,
X^{ik}_{p,\,p^{\prime}}(\Omega,\,j)\,-\,
\sum\limits_{h,\,h^{\prime}} X^{ik}_{p,\,h}
[X^{ik}_{h,\,h^{\prime}}(\Omega,\,j)]^{-1}
X^{ik}_{h^{\prime},\,p^{\prime}}(\Omega,\,j)
\label{eq62}
\end{eqnarray}
and
\begin{eqnarray}
h^{ik}_{p,\,p^{\prime}}(\Omega,\,j)\,\equiv\,
n^{ik}_{p,\,p^{\prime}}(\Omega,\,j)
h^{ik}(\Omega,\,j)\,+\,\sum\limits_{\alpha\sigma,\,\gamma\sigma^{\prime}}
\omega^{ik}_{p,\,\alpha\sigma}(\Omega,\,j)
{\tilde\Gamma}^{ik}_{\alpha\sigma,\,\gamma\sigma^{\prime}}(\Omega,\,j)
{\tilde\omega}^{ki}_{\gamma\sigma^{\prime},\,p^{\prime}}(\Omega,\,j)
\,,
\label{eq63}
\end{eqnarray}
respectively. Furthermore,
\begin{eqnarray}
\omega^{ik}_{p,\,\alpha\sigma}(\Omega,\,j)\,\equiv\,
\sum\limits_{\beta,\,\sigma^{\prime}}
D^i_{\beta\sigma^{\prime},\,p}[{\bf 1}\,-\,
{\tilde\rho}^{ki}(\Omega,\,j)]_{\beta\sigma^{\prime},\,\alpha\sigma}
\,,
\label{eq64}
\end{eqnarray}
and all the other functions are again defined in sec.~\ref{ssec22}. 
For each possible linear momentum quantum number $\xi_{+1}\;$
the eqs.~(\ref{eq58}) and (\ref{eq59}) yield 
$2m\cdot (2N-N_e)$ solutions $\tilde p$ with $S=1/2$ and
energies $E_{\tilde p}$.

\noindent
Now the hole--spectral functions can be calculated. 
The spectral function for hole states, 
$S_{\tilde h}(k,\,\epsilon_{\tilde h})$, is defined by
\begin{eqnarray}
\sqrt{S_{\tilde h}(k,\,\epsilon_{\tilde h})}
&\equiv& \langle \tilde h\,;\,N_e-1\,\xi_{-1}\,S=1/2\vert\vert\,
\co_{\xi_0-\xi_{-1}}\,\vert\vert D^1;\,N_e\,\xi_0\,S=0\rangle\cr
&=&-\sqrt{2} n_0^{-1/2}\sum\limits_{i=1}^m\sum\limits_{h=1}^{N_e}
\sum\limits_{\sigma=-1/2}^{1/2}\,f^{{1/2\,\xi_{-1}}^*}_{i h\sigma,\,{\tilde h}}
\,\cdot\cr
&&\qquad\cdot\,\frac{1}{N}\sum\limits_{j=1}^N\exp\left\{-i\frac{2\pi}{N}
\xi_{-1}\,j\right\}\frac{1}{8\pi^2}\sum\limits_{\sigma^{\prime}=-1/2}^{1/2}
\int d\Omega D^{{1/2}^*}_{\sigma,\,\sigma^{\prime}}(\Omega)\,\cdot \cr
&&\qquad\cdot\,\left[\sum\limits_{h^{\prime}=1}^{N_e} (-1)^{1/2-\sigma^{\prime}}
D^{1^*}_{\xi_0-\xi_{-1}\,-\sigma^{\prime},\,h^{\prime}}
(X^{i1})^{-1}_{h^{\prime},\,h}(\Omega,\,j)\right]n^{i1}(\Omega,\,j)\,,
\label{eq65}
\end{eqnarray}
where $k=(2\pi/N)\xi_{-1}$ and $\epsilon_{\tilde h}=E_0-E_{\tilde h}$.
Similarly, the particle spectral function can be obtained from
\begin{eqnarray}
\sqrt{S_{\tilde p}(k,\,\epsilon_{\tilde p})}
&\equiv&\langle \tilde p\,;\,N_e+1\,\xi_{+1}\,S=1/2\vert\vert\,
\co^{\dagger}_{\xi_{+1}-\xi_0}\,\vert\vert D^1;\,N_e\,\xi_0\,S=0\rangle\cr
&=&-\sqrt{2} n_0^{-1/2}\sum\limits_{i=1}^m\sum\limits_{p=N_e+1}^{2N}
\sum\limits_{\sigma=-1/2}^{1/2}\,g^{{1/2\,\xi_{+1}}^*}_{i p\sigma,\,{\tilde p}}
\,\cdot\cr
&&\qquad\cdot\,\frac{1}{N}\sum\limits_{j=1}^N\exp\left\{-i\frac{2\pi}{N}
\xi_{+1}\,j\right\}\frac{1}{8\pi^2}\sum\limits_{\sigma^{\prime}=-1/2}^{1/2}
\int d\Omega D^{{1/2}^*}_{\sigma,\,\sigma^{\prime}}(\Omega)\,\cdot\cr
&&\qquad\cdot\,\left[\sum\limits_{p^{\prime}=N_e+1}^{2N}
n^{i1}_{p,\,p^{\prime}}(\Omega,\,j)
D^{1}_{\xi_{+1}-\xi_0\,\sigma^{\prime},\,p^{\prime}}\right]
n^{i1}(\Omega,\,j)\,,
\label{eq66}
\end{eqnarray}
where now $k=(2\pi/N)\xi_{+1}$ and $\epsilon_{\tilde p}=E_{\tilde p}-E_0$.

\noindent
The sum
\begin{eqnarray}
\sum\limits_{{\tilde h}=1}^{2m\cdot N_e}\,
S_{\tilde h}(k,\,\epsilon_{\tilde h})\equiv\,n(k=(2\pi/N)\xi_{-1})
\label{eq67}
\end{eqnarray}
gives the occupation number of the basis state~(\ref{eq3}) in the ground state
of the $N_e$--electron system. Here instead of using $\xi_{-1}=0,\dots,N-1$, 
we use the equivalent representation $\xi_{-1}=-N/2+1,\dots,N/2$.

\noindent
Furthermore, for plotting the  spectral functions
versus the linear momentum $k$ and excitation energy
$\omega=\epsilon_{\tilde h}$ (or $=\epsilon_{\tilde p}$)
it is useful to introduce some artificial width for each state in
order to obtain continuous functions of the excitation energy. For this
purpose we use a Lorentzian shape with fixed width of 0.05 $t$ for each
hole (or particle) state. This also simplifies the representation
of the density of states
\begin{eqnarray}
N(\omega)\,\equiv\,\sum\limits_{\alpha=-N/2+1}^{N/2}\,\left[
S_{(\tilde h)}((2\pi/N)\alpha,\,\omega)\,+\,
S_{(\tilde p)}((2\pi/N)\alpha,\,\omega)\,\right]
\,.
\label{eq68}
\end{eqnarray}
where the indices ${\tilde h}$ and ${\tilde p}$ have been put here in
parentheses, since they are absorbed now in the continuous variable
$\omega$.

In order to determine the hole spectral functions from an exact (Lanczos) 
solution,
we  start with the exact ground-state of the system with $N_e$ electrons,
$\vert N_e\rangle_0$, which is represented by a linear combination of the basic
configurations and apply the annihilation for an electron with spin projection
$\sigma$ and momentum $\alpha = \xi_0 - \xi_{-1}$
\begin{equation}
\co_{\alpha,\sigma} \vert N_e\rangle_0 = \eta_{\alpha}
\vert N_e-1, \xi_{-1}, S=1/2\rangle_1\,.\label{eq:exac1}
\end{equation}
The resulting state $\vert N_e-1, \xi_{-1}, S=1/2\rangle_1$ is a state with
$N_e-1$ electrons and well defined quantum numbers for the momentum ($\xi_{-1}$) and
spin. The constant $\eta_{\alpha}$ has been introduced to normalize the state.
This state is in general not an eigenstate of the Hamiltonian, but it can be
used as a starting point to generate further states $\vert N_e-1, \xi_{-1},
S=1/2\rangle_i$ with the same quantum numbers by means of the Lanczos
method\cite{bagel}. A diagonalisation of the Hamiltonian in the space which is
generated by these basis states $\vert N_e-1, \xi_{-1},S=1/2\rangle_i$ for $i=1
\dots \nu$ will lead to $\nu$ eigenvalues $\varepsilon_\alpha^i$ and eigenstates
$\vert \varepsilon_\alpha^i\rangle$, which converge
with increasing $\nu$ to the exact solutions for the system with $N_e-1$
electrons. The spectral function for the hole states can now be expressed
as
\begin{equation}
S_h((2\pi/N)\alpha,\omega) = \eta_{\alpha}^2 \sum_{i=1}^{\nu}
\left\vert\langle  \varepsilon_\alpha^i\vert N_e-1,
\xi_{-1},S=1/2\rangle_0\right\vert^2 \delta\left(\omega -
\varepsilon_\alpha^i\right)\,.\label{eq:exac2}
\end{equation}
This means that the occupation probability $n((2\pi/N)\alpha)$ is 
determined by the square of the normalisation constant $\eta_{\alpha}$ defined
in eq.~(\ref{eq:exac1}) and the overlap of the states $\vert \varepsilon_\alpha^i
\rangle$ with the starting vector for the Lanczos iteration $\vert N_e-1, \xi_{-1}, 
S=1/2\rangle_1$ defines the spectral strength at the energy $\omega = 
\varepsilon_\alpha^i$. The result for the spectral function converges very
rapidly with the number of iterations $\nu$. A corresponding procedure starting
with a state
\begin{equation}
\co^\dagger_{\alpha,\sigma} \vert N_e\rangle_0  \,,\label{eq:exac3}
\end{equation} 
can be used to determine the spectral function for the particle states.

\section{RESULTS AND DISCUSSION\label{sec3}}

\begin{figure}[htb]
%\center
\includegraphics[scale=0.4,angle=270]{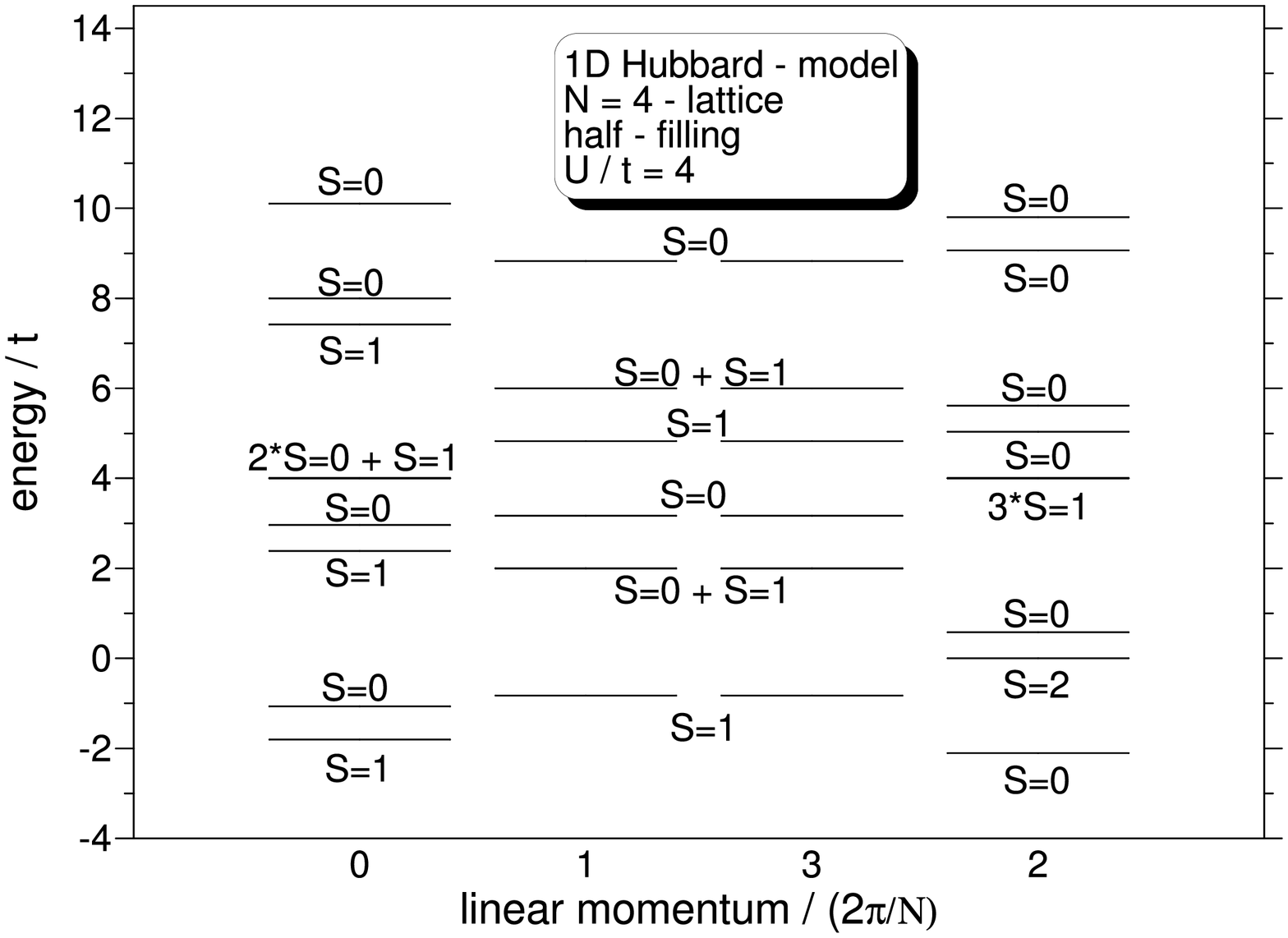}
\caption{Energy spectrum of the half--filled $N=4$--lattice as obtained
with interaction strength $U/t=4$ either by complete diagonalization or by
the LMSPHF--approach out of sec.~\ref{ssec22}. In the latter case each of the
states is represented by a single linear momentum-- and spin--projected
Slater--determinant.}
\label{fig1}
\end{figure}

First, we checked the LMSPHF--approach discussed in sec.~\ref{ssec22} for
half--filled lattices associated to $N=2$, $N=4$ and $N=6$ sites, respectively. Simple
counting of the number of variational variables (see tab.~\ref{tab1}) shows that
for these lattices the approach is exact. Each state can be represented by a
single spin-- and linear momentum--projected HF--configuration for any value
of $U/t$. This has been tested explicitely for the 36 different states of the
half--filled $N=4$ lattice as well as for the 5 lowest states of each possible
linear momentum quantum number with spins $S=0$, $S=1$ and $S=2$ in the
half--filled $N=6$ grid. As an illustrative example we present in
Fig.~\ref{fig1} the energy spectrum for the half-filled $N=4$ lattice as obtained
for $U/t=4$. Each of these states can be represented by a single projected
LMSPHF--determinant. The figure also shows
nicely the degeneracy of the states with $\xi=1$ and $\xi=3$ (or -1)
discussed in sec.~\ref{ssec21}. 
As expected, the ground states have all total spin
$S=0$, however, while for the lattices with $N=2$ and $N=6$ the linear
momentum quantum number $\xi=0$ is obtained, the $N=4$ ground state
has $\xi=N/2=2$. 

\begin{table}[htb]
\begin{tabular}{c|ccccc}
\hline
$N$        & 4 & 6 & 12                & 14                & 30 \\ \hline 
Hilbert restricted space  & $~\sim 10$&$\sim 70$&$\sim 2\times 10^4$&$\sim
2\times 10^5$&$\sim  10^{14}$   \\
variationnal parameters & 32& 72& 288               & 392               & 1800   \\
\hline
\end{tabular}
\caption{Comparison of the dimension of the Hilbert restricted space
  (taking account of the symmetries) for a given
  half-filled lattice with $N$ sites with the number of parameters
  involved in the LMSPHF-approach.}
\label{tab1}
\end{table}

\noindent
This feature persists, if half--filled lattices with larger $N$ are
considered~: For $N/2$ being an odd integer the ground state 
has $S=0$ and $\xi=0$, while for $N/2$ being an even integer $S=0$ and
$\xi=N/2$ is obtained. 

\noindent
This effect can be understood by a simple spin--correlation. Assume for the
moment small interaction strength $U$. Then, if $N/2$ is odd, the
configuration with the lowest energy is obviously the determinant in which
the single particle states~(\ref{eq3}) with $\alpha=0,\dots,\,\pm (N-2)/4$ are all
filled, each by two electrons with opposite spin--directions. The resulting
total spin is then $S=0$ and the linear momentum quantum number $\xi=0$. For
$N/2$ being even, however, the situation is different. Here, the last
two electrons have to be distributed over the two degenerate states
with $\alpha=\pm N/4$. 

\noindent
It can be shown that for two electrons with opposite spin
directions occupying two 
different basis orbits~(\ref{eq3}) always the $S=1$ component is energetically
favored with respect to the $S=0$ component. The spin--correlation explains
why half--filled lattices with $N/2$ being odd has the lowest excited
state associated with total spin $S=1$ and linear momentum 
quantum number $\xi=N/2$. Assuming again small interaction
strength $U$, the energetically lowest configurations are obtained by
promoting one of the four electrons from the last occupied orbits
($\alpha=\pm (N-2)/4$) to the first unoccupied orbits ($\alpha=\pm (N+2)/4$) 
in the above mentioned $(S=0,\,\xi=0)$--ground state configuration
for odd $N/2$. There are 8 degenerate determinants of this type, 
4 of them with $\xi=N/2$, 2 with $\xi=1$ and 2 with $\xi=-1$ (or, equivalent
$\xi=N-1$). The expectation value of the Hamiltonian for each of these
determinants will be denoted by $b$. The $\xi=\pm 1$ configurations yield
two degenerate $S=1$ states at energy $b-U/N$ and two degenerate $S=0$ states
at energy $b+U/N$. The four $\xi=N/2$ determinants, however, can be coupled to
two $S=1$ configurations both with energy $b-U/N$ and an interaction of
$-U/N$ between them and to two $S=0$ configurations both with energy
$b+U/N$ and an interaction of $U/N$ in between. Thus we get here
one $S=1$ solution with energy $b-2U/N$, an $S=1$ and an $S=0$ solution
both having energy $b$, and one $S=0$ state with energy $b+2U/N$.
Consequently, here the lowest $(S=1,\,\xi=N/2)$--solution will become
the first excited state. Again the relative splitting in between the
various states will even increase with increasing interaction. Also
this expectation is confirmed by the results obtained for the half--filled
$N=6$--, $N=14$-- and $N=30$--lattices in the present work.

\subsection{Half-filled $N$=12 lattice}

\begin{figure}[htb]
%\center
\includegraphics[scale=0.4,angle=270]{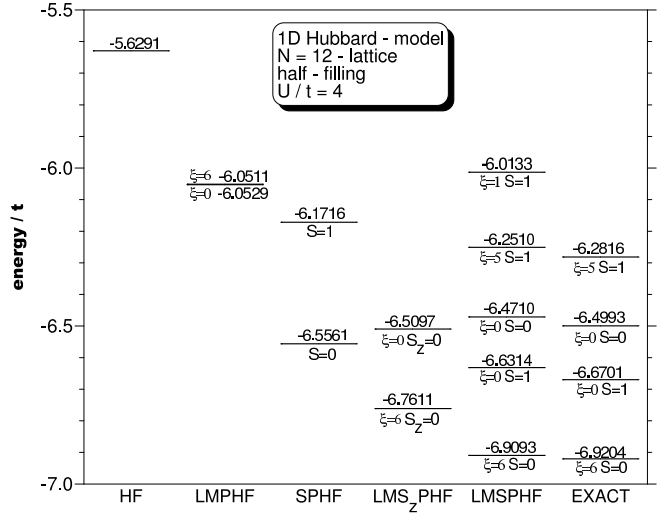}
\caption{Energies for some states of the half--filled $N=12$--lattice
as obtained using the interaction strength $U/t=4$ by various approaches. 
The different columns refer to a simple unprojected HF--calculation (HF),
to HF with only linear momentum-- (LMPHF), or only spin--projection (SPHF)
before the variation and to HF with restoration of the total linear momentum
and only the 3--component of the total spin (LMS$_z$PHF). Finally, the
energies of the lowest few states obtained with linear momentum-- and full
spin--projection before the variation as described in sec.\ref{ssec22} (LMSPHF)
are compared with those resulting from a complete diagonalization (EXACT).}
\label{fig2}
\end{figure}

Let us now consider the half--filled $N=12$ lattice. Here in the usual
Lanczos approach (all $\Sigma=0$ determinants) already 853 776 configurations
have to be treated, and even using all the symmetries, for $S=0$ and the 12
possible  $\xi$--values still between 18 840 and 18 916, for $S=1$ and the
possible $\xi$--values between 31 833 and 31 872 configurations have to be
accounted for.
Fig.~\ref{fig2} displays the energies of the first excited states for
various methods using an interaction strength of $U/t=4$. 
In the
rightmost column (EXACT) the results of the Lanczos diagonalisation
for the lowest 4 states are presented, the other columns refer to different 
variational approaches. The leftmost one (HF) gives the Hartree-Fock result 
obtained by using the
determinant (12) as test wave function and not caring about spin-- and
linear momentum. The next two columns have been obtained by HF calculations
with linear momentum projection before the variation only (LMPHF)
but still not restoring the total spin and with spin--projection
before the variation only (SPHF) but not restoring the total linear momentum.
The next column (LMS$_z$PH) results from variational calculations with 
projection on good linear momentum and z--projection of the spin $S_z=0$.
Finally, the second but last column (LMSPHF) presents the results 
obtained for the lowest 5 states with linear momentum-- and full
spin--projection before the variation as described in sec.~\ref{ssec22}. As can
be seen, the LMSPHF--approach reproduces the energies of the exact solutions
not only for the ground but also for the lowest excited states very well.
The deviations vary only between 0.16 and 0.58 percent. 
This is remarkable since the number of variational parameters (288 for
$S=0$ in this case) is significantly smaller than the dimension of the space
with good symmetries, which is 18840 in this example.

\noindent
As expected from the above arguments based on spin--correlation
the ground state has $S=0$ and $\xi=N/2=6$, while for the
quantum numbers of the first excited state $S=1$ and $\xi=0$ are obtained.
It is furthermore obvious, that both, linear momentum and spin have to be
restored simultaneously. All results obtained by performing none or only
part of the corresponding projections fail to reproduce the exact spectrum.

\begin{figure}[htb]
%\center
\includegraphics[scale=0.5,angle=270]{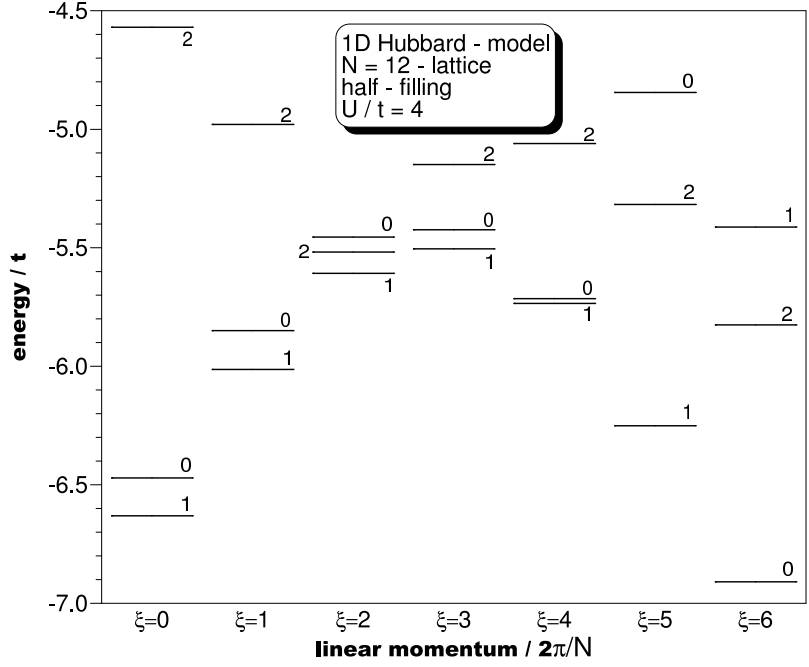}
\caption{Again for the half--filled $N=12$--lattice and $U/t=4$,
the energy spectra for the lowest $S=0$, $S=1$ and $S=2$ states
as obtained with the LMSPHF--method for the various linear momentum
quantum numbers $\xi$ are displayed. Because of the degeneracy of the
$\xi=12-i$ with the $\xi=i$ spectra for $i=1,\dots,5$, only the spectra
for $\xi=0,\dots\,6$ are presented.}
\label{fig3}
\end{figure}

\noindent
Fig.~\ref{fig3} presents the energies of lowest $S=0$, $S=1$ and $S=2$ states
obtained with the LMSPHF--approach for the various linear momentum
quantum numbers $\xi$. Note, that the $\xi=1$--results are degenerate
with those for $\xi=11$, those for $\xi=2$ with those for $\xi=10$, etc.,
so that only the results for $\xi=0,\dots\,6$ are shown. Except for
the $(S=0,\,\xi=N/2)$--ground state, the lowest states for all other
linear momentum quantum numbers have total spin $S=1$. This supports
again the importance of the above discussed spin--correlation.

\begin{figure}[htb]
%\center
\includegraphics[scale=0.5,angle=270]{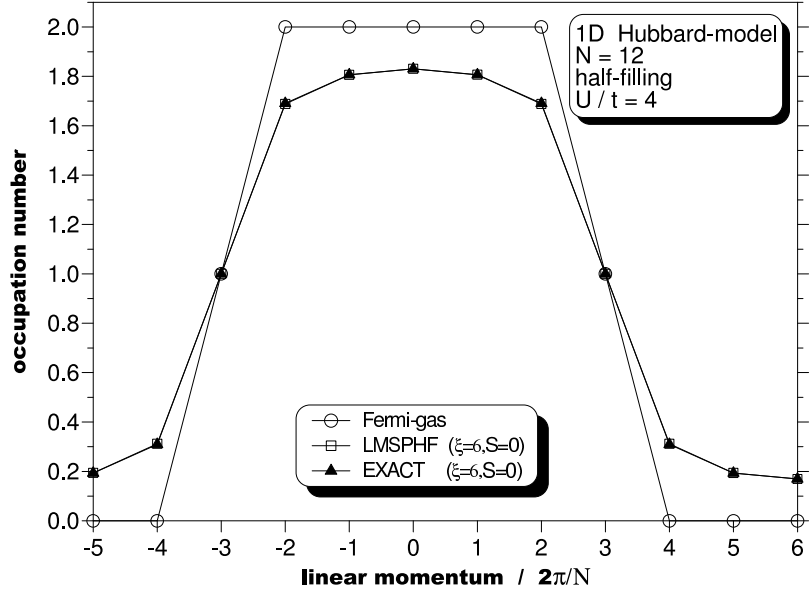}
\caption{For the half--filled $N=12$--lattice and $U/t=4$ the occupation
numbers (eq.~(\ref{eq66})) of the various basis states in the
$(S=0,\,\xi=6)$--LMSPHF--ground state are compared to those obtained by a
complete diagonalization (EXACT) and to those expected for a non--interacting
Fermi--gas. Identical occupation numbers are obtained for using only the
lowest ($m=1$) or the lowest five ($m=5$) determinants in the calculation of
the hole--spectral functions.}
\label{fig4}
\end{figure}

\noindent
In fig.~\ref{fig4}, we represent the occupation numbers of the various basis 
orbitals~(\ref{eq3}) in the LMSPHF--ground state obtained via
eq.~(\ref{eq67}) and we compare it with those
resulting from the one--body density of the exact ground state and with
those expected for a non--interacting Fermi--gas. Again, excellent agreement
of the LMSPHF--results with the exact solution is obtained. It should be
stressed, that the occupation numbers obtained via eq.~(\ref{eq67}) are
not affected by the number of determinants $m$ used for the calculation
of the hole--spectral functions. Identical numbers are obtained, no matter
whether only the ground state determinant $\vert D^1\rangle$ ($m=1$)
or, e.g., all the $m=5$ determinants corresponding to the five lowest
LMSPHF--solutions Fig.~\ref{fig2} are included in the calculation.

\begin{figure}[htb]
%\center
\includegraphics[scale=0.5,angle=270]{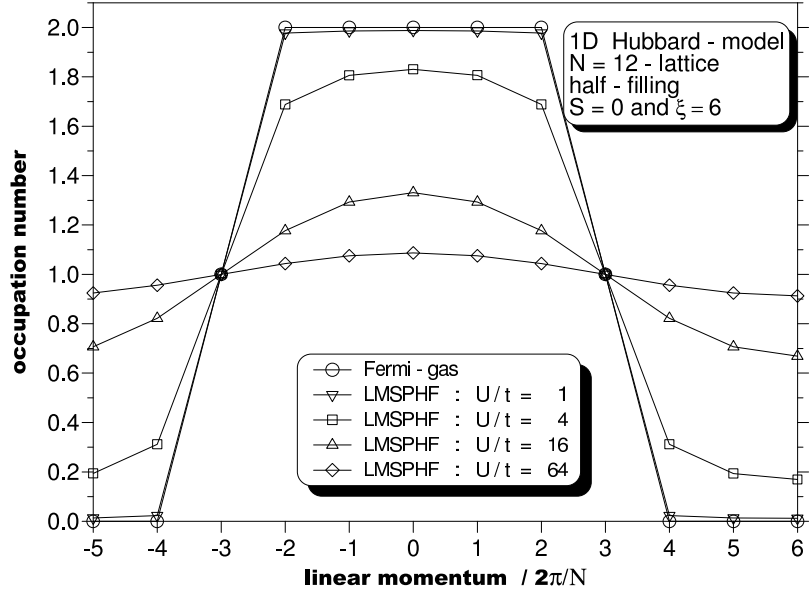}
\caption{Same as in Fig.~\ref{fig4}, but for various interaction strengths
$U/t$. Since in all cases the exact and LMSPHF-results cannot be
distinguished, only the latter are displayed.}
\label{fig5}
\end{figure}

\noindent
In fig.~\ref{fig5}, we compare the occupation numbers of the basis orbitals in the 
LMSPHF--($S=0,\,\xi=6$)--ground states obtained for various strength
parameters $U/t$ of the interaction. Since in all cases the LMSPHF-- and
exact occupation numbers cannot be distinguished, we have plotted
  only the former. As expected, the correlations (i.e, the deviation from
the Fermi--gas values) grow with increasing interaction. Already at
$U/t=64$ the result looks rather similar to the equal distribution
of the anti-ferromagnetic limit expected for $U/t\,\to\,\infty$.

\begin{figure}[htb]
%\center
\includegraphics[scale=0.5,angle=270]{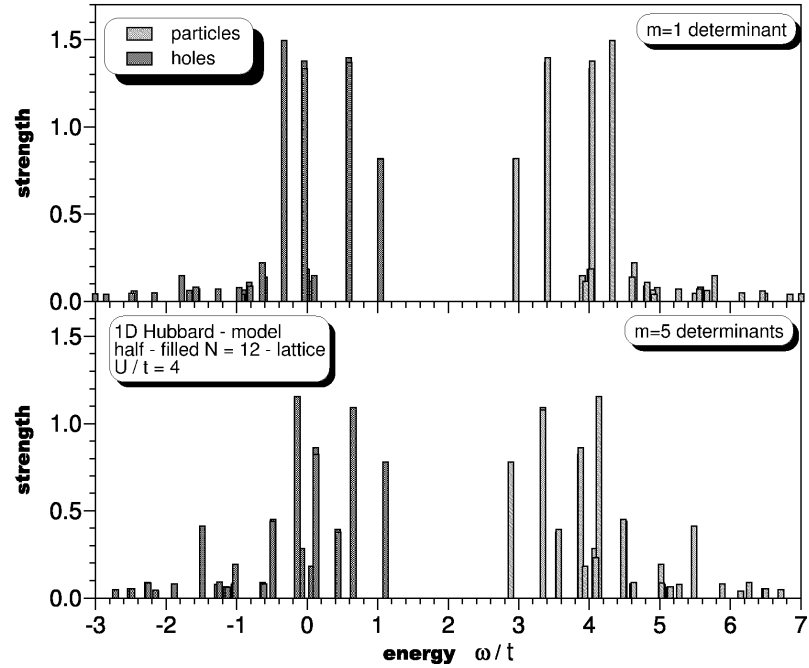}
\caption{Again for the half--filled $N=12$--lattice and $U/t=4$,
the hole-- (eq.~(\ref{eq65})) and particle-- (eq.~(\ref{eq66})) spectral
functions are plotted for all possible values of $\xi_{\pm 1}$ as
functions of the energy $\omega$ ($=\epsilon_{\tilde h}$ and
$=\epsilon_{\tilde p}$, respectively). The upper part of the figure
has been obtained by using only the ground state determinant 
$\vert D^1\rangle$ in the calculation for the spectral functions ($m=1$).
Using instead  all the $m=5$ determinants corresponding to the lowest five
states out of Fig.~\ref{fig2} in these calculations, one obtains the results
displayed in the lower part of the figure.}
\label{fig6}
\end{figure}

\noindent
In fig.~\ref{fig6} we represent  the
hole-- (eq.~(\ref{eq65})) and particle-- (eq.~(\ref{eq66}))
spectral functions versus the excitation energy $\omega$
(in units of $t$). Here, $\omega=\epsilon_{\tilde h}$ for the hole-- and
$\omega=\epsilon_{\tilde p}$ for the particle--states, respectively. 
We compared the results obtained by using only one determinant ($m=1$, upper half of
the figure) with those resulting from using  the HF--transformations
obtained for all the $m=5$ lowest LMSPHF--states presented in Fig.~\ref{fig2}.
As expected for the half--filled
lattice, particle-- and hole--states are nicely symmetric around the
Fermi--energy $\omega_F\,=\,U/t/2$. They are separated by the so--called 
``Hubbard--gap'' of about $U/t/2$. Using $m=5$ determinants one obtains 
 120 hole and 120 particle states for each possible linear
momentum quantum number, while we obtain only 24 states for $m=1$ 
for each $\xi_{\pm 1}$--value. Consequently, the strength is more
spread for the $m=5$--approximation with respect to the more
restricted $m=1$--approximation. 

\begin{figure}[htb]
%\center
\includegraphics[scale=0.5,angle=270]{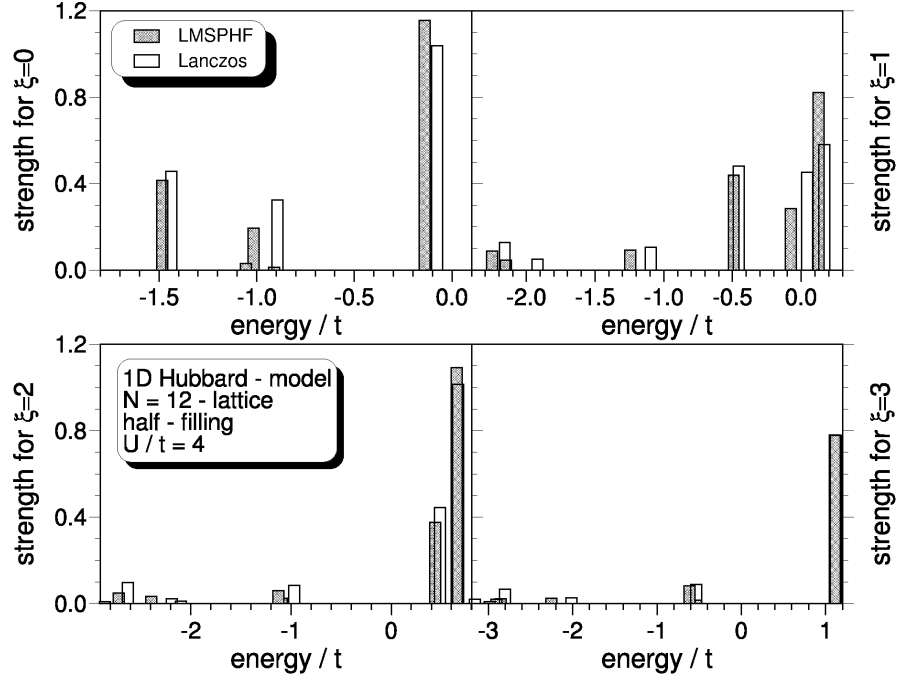}
\caption{Again for the half--filled $N=12$--lattice and $U/t=4$,
the  hole spectral functions for $\xi=0,\dots,3$
as obtained with $m=5$ determinants are plotted versus the excitation
energy and compared with the ``exact'' results computed with the
Lanczos--approach as discussed in the text.}
\label{fig7}
\end{figure}

\noindent
Finally, in Fig.~\ref{fig7} we compare the results for the strength
defined by eq.~(\ref{eq66}) and obtained with $m=5$ determinants for
$S=1/2$ and $\xi=0,\dots,3$ with those resulting from the exact
calculation as described in (\ref{eq:exac1}) - (\ref{eq:exac3}).
As can be seen, the agreement for both the excitation energies as well as
for the splitting of the strength is again very good.

\subsection{Doped lattice}

\noindent
Let us now consider a system of only 10 electrons in the $N=12$--lattice,
again using $U/t=4$ as interaction strength. 
$N_e/2$ is odd, then all the basis orbits~(\ref{eq3})
with $\alpha=0,\,\pm 1$ and $\pm 2$ can be filled, each with two electrons
with opposite spin directions. Thus, from spin--correlation argument, 
one expects an $(S=0,\,\xi=0)$--ground
state. The lowest excited states can then be obtained by promoting one
of the four electrons from the $\alpha=\pm 2$--orbitals to the
$\alpha=\pm 3$ orbits. There are again 8 degenerate determinants of this
type~: 2 with $\xi=1$, 2 with $\xi=-1$, 2 with $\xi=5$ and 2 with
$\xi=-5$. Calling $\mathcal{C}$ the expectation value of the
Hamiltonian~(\ref{eq5}) within each of these determinants, one obtains
four degenerate $S=1$ states at energy $\mathcal{C}-U/N$ 
and four degenerate $S=0$ states at energy $\mathcal{C}+U/N$,
respectively. The corresponding linear momentum quantum
numbers are in both cases $\xi=1,-1,5$ and $-5$. Thus one expects
the lowest excited states to have $S=1$ and $\xi=\pm 1,\,\pm 5$.

\begin{figure}[htb]
%\center
\includegraphics[scale=0.5,angle=270]{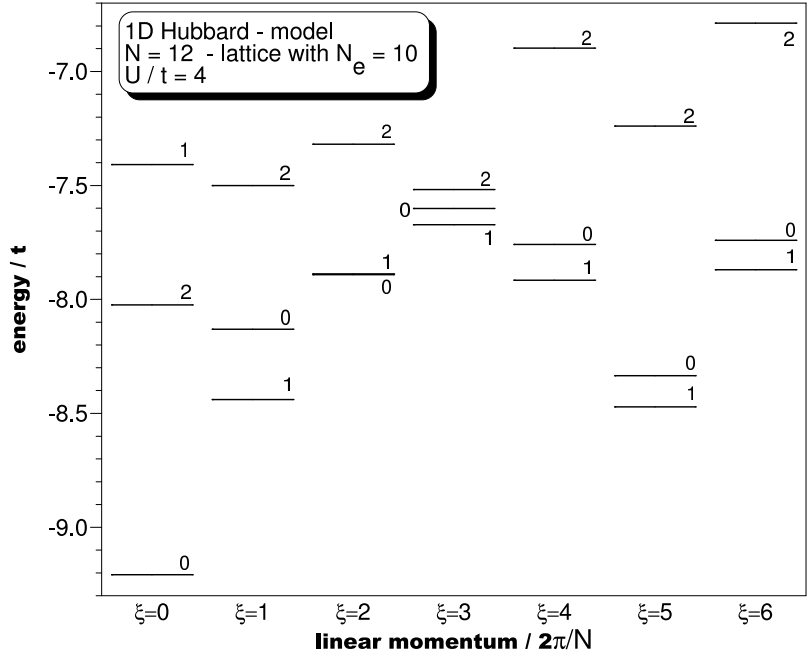}
\caption{Same as in Fig.~\ref{fig3}, but for only 10 electrons in the
$N=12$--lattice.}
\label{fig8}
\end{figure}

\noindent
Fig.~\ref{fig8}  presents the energies of lowest $S=0$, $S=1$ and $S=2$ states
of the 10--electron system on the $N=12$--lattice obtained with the
LMSPHF--approach for the various linear momentum quantum numbers 
$\xi=0,\dots\,6$. As in Fig.~\ref{fig3}, the spectra for $\xi=-1,\dots,-5$,
which are degenerate with those for $\xi=1,\dots\,5$ have not been
plotted. As expected, we obtain an $(S=0,\,\xi=0)$--ground state,
and the lowest four excited states have all $S=1$ and linear momentum
quantum numbers of $\xi=\pm 1$ and $\xi=\pm 5$, respectively.

\begin{figure}[htb]
%\center
\includegraphics[scale=0.5,angle=270]{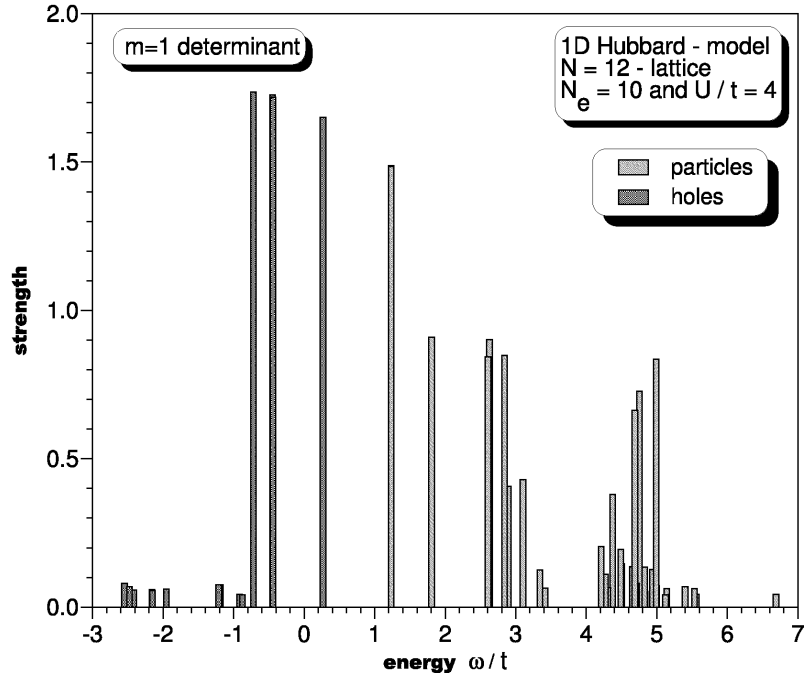}
\caption{Same as in Fig.~\ref{fig6}, but for only 10 electrons in the
$N=12$--lattice. Here only the results obtained with one determinant
($m=1$) are presented.}
\label{fig9}
\end{figure}

\noindent
Obviously, the nice particle--hole symmetry in the spectral functions 
of the half--filled 
$N=12$--lattice is destroyed, if only 10 electrons in this lattice are
considered. Using the simplest ($m=1$) approximation for the calculation
of the $S=1/2$--wave functions for the 9-- and 11--electron systems, one
obtains here for each possible linear momentum 20 one--hole-- and 28
one--particle--states. This is reflected in the 
hole-- and particle--spectral functions, which are shown in Fig.~\ref{fig9}
as a function of the excitation energy. Particle-- and hole--strengths are now
distributed asymmetrically around the Fermi--energy and furthermore,
the Hubbard--gap obtained for the half--filled lattice has vanished.

\subsection{$N$=14 lattice}

\noindent
We shall now consider the half--filled $N=14$--lattice, again using
$U/t=4$. Here, in the usual Lanczos--approach (all determinants
with $\Sigma=0$) already 11 778 624 determinants have to be treated.
Consequently, for this example we could obtain only the exact ground
and first exited state within about one week of computer time. Even
if all the symmetries would be used, for $S=0$ and the 14 possible  
linear momentum quantum numbers still between 197 099 and 197 276,
for $S=1$ and the possible $\xi$--values between 357 770  and 357 945
configurations have to be accounted for.

\begin{figure}[htb]
%\center
\includegraphics[scale=0.5,angle=270]{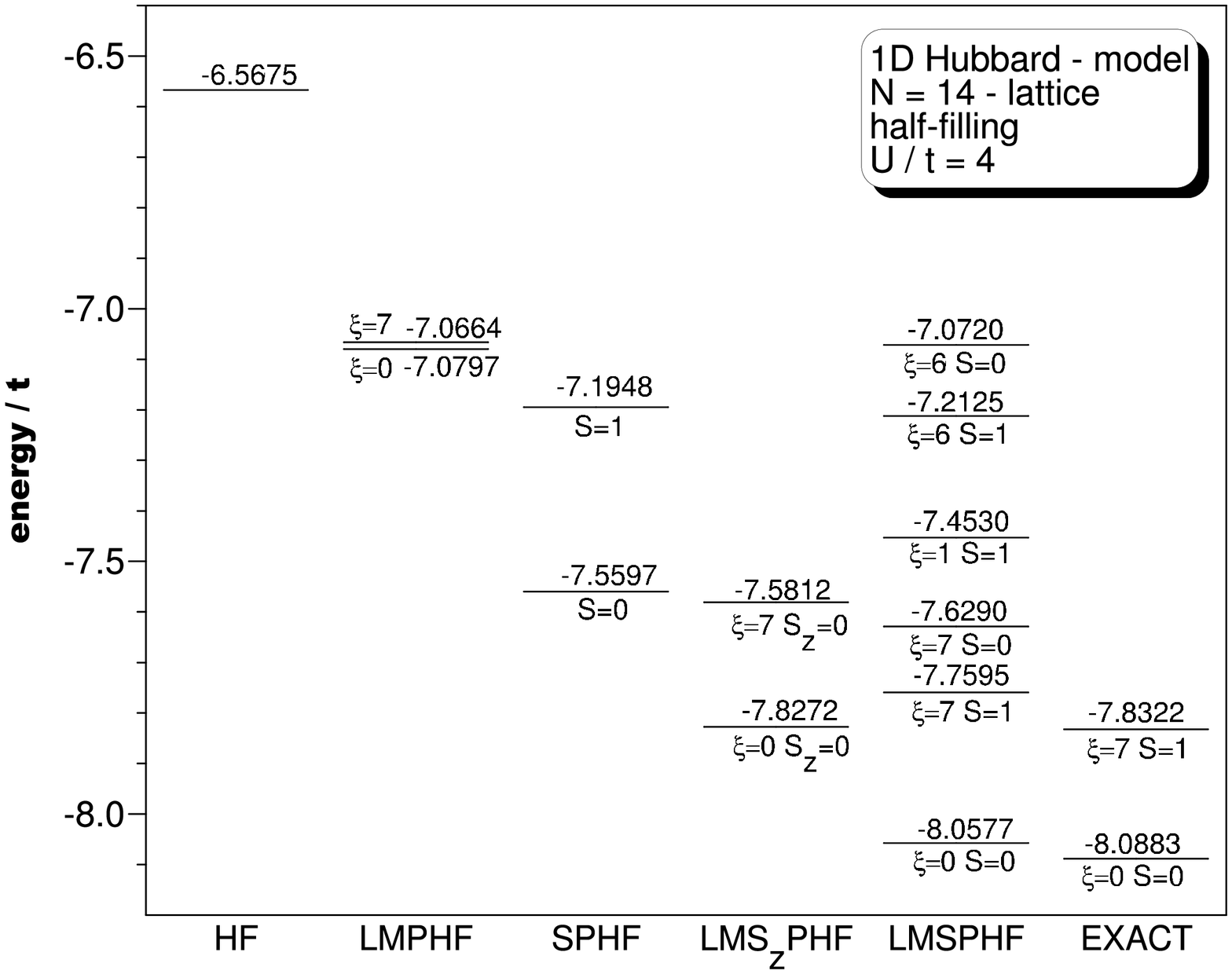}
\caption{Same as in Fig.~\ref{fig2}, but for the half--filled $N=14$--lattice.}
\label{fig10}
\end{figure}

\noindent
In fig.~\ref{fig10}, we display the results for the lowest states obtained
with different variational methods and compare them with the exact
results. The nomenclature is the same as in Fig.~\ref{fig2}. Again it is seen,
that all results obtained by performing none or only part of the
symmetry--projections fail to reproduce the exact spectrum. Only
if linear-- momentum and full spin--projection are both performed
simultaneously before the variation as done in the LMSPHF--approach
the exact data can be reproduced. The deviation of the LMSPHF--energies
from the latter amount here to 0.38 percent for the ground and 0.93
percent for the first excited state, respectively. Thus (as expected
because of the considerably larger dimensions) they are larger than
in the half--filled $N=12$--example. It should be stressed, however,
that in the LMSPHF--approach each of these states are represented by
only a single symmetry--projected configuration. Correlating these
solutions by additional determinants could, obviously, still improve
the agreement. As expected for a half--filled lattice with $N/2$ being odd,
the ground state has total spin $S=0$ and linear momentum quantum number
$\xi=0$, while for the first excited state $S=1$ and $\xi=N/2=7$ is
obtained.

\begin{figure}[htb]
%\center
\includegraphics[scale=0.5,angle=270]{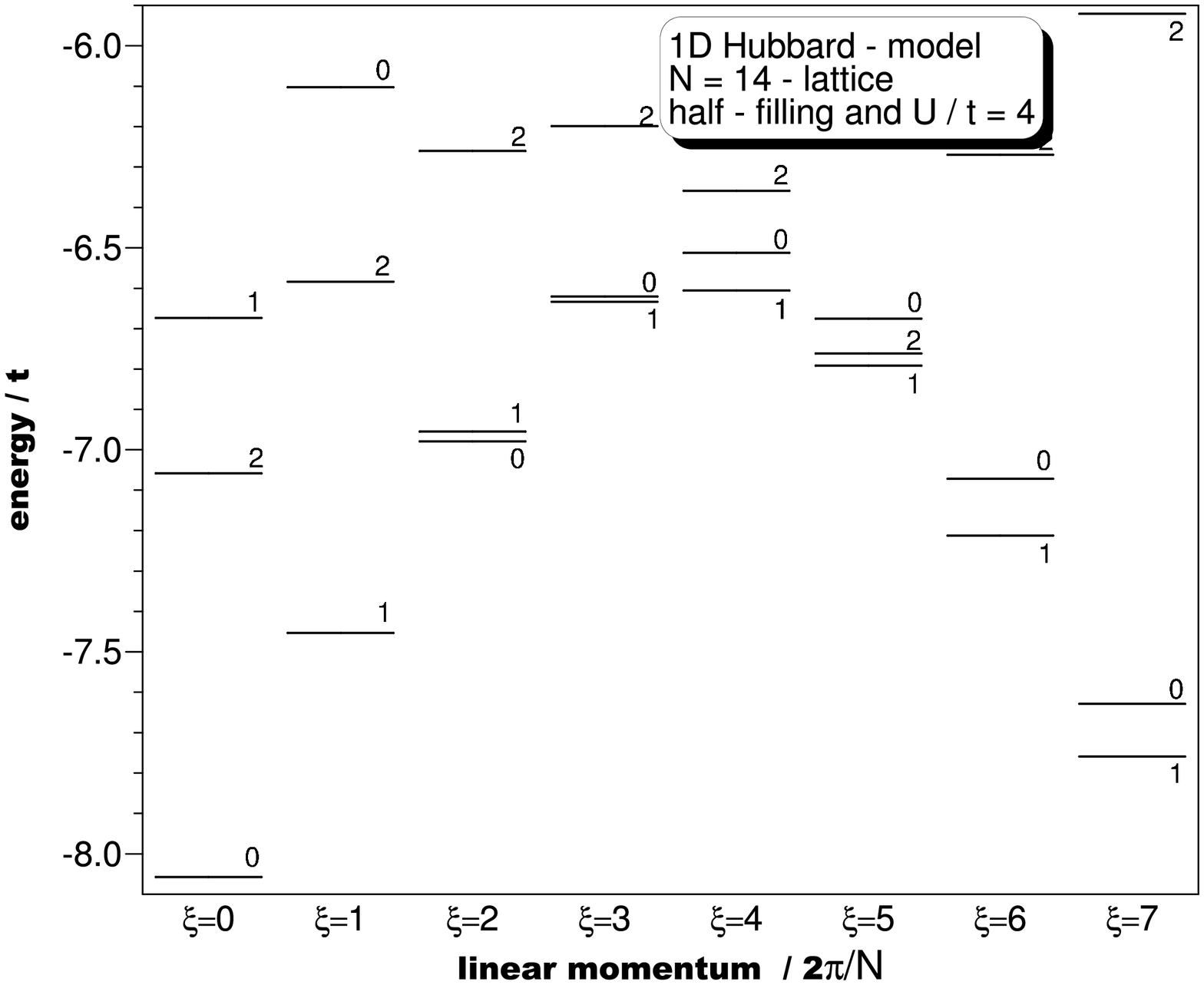}
\caption{Same as in Fig.~\ref{fig3}, but for the half--filled $N=14$--lattice.}
\label{fig11}
\end{figure}

\noindent
In fig.~\ref{fig11} we represent the energies of the lowest $S=0$, $S=1$ and $S=2$ states
obtained with the LMSPHF--approach for the various linear momentum
quantum numbers $\xi$. Again, the spectra for $\xi=N/2+1,\dots,(N-1)$
are degenerate with those obtained for $\xi=1,\dots\,N/2-1$, respectively,
and are hence not displayed. Except for the $(S=0,\,\xi=0)$--ground 
states and the $\xi=2$ case, where the lowest $S=0$ and $S=1$ states are
almost degenerate, the lowest states for all the other linear momentum
quantum numbers have always total spin $S=1$. This supports 
the above discussed spin--correlation favoring the $S=1$--excitations.

\begin{figure}[htb]
%\center
\includegraphics[scale=0.5]{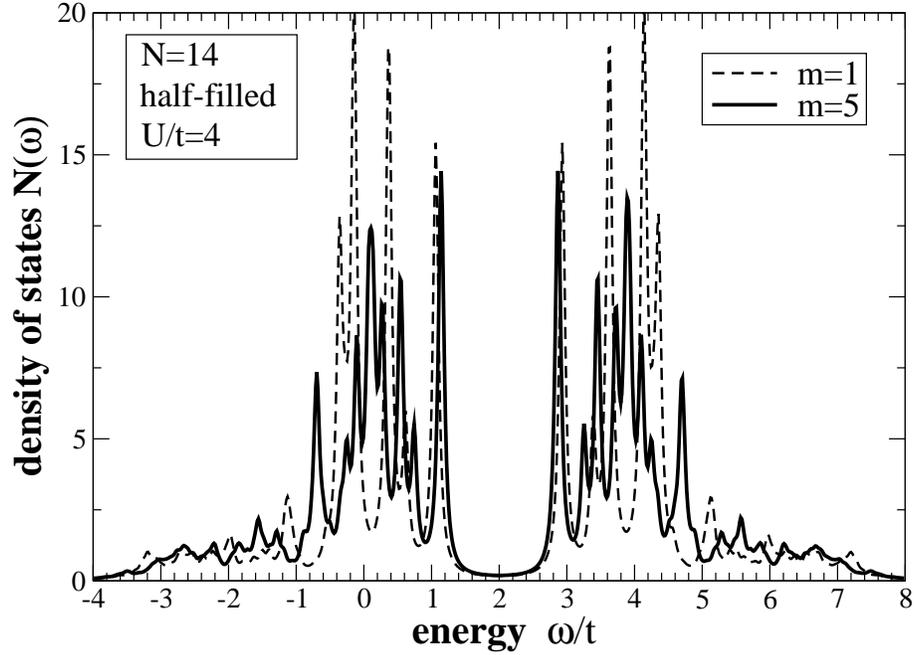}
\caption{Again for the half--filled $N=14$--lattice, the sum of the
 hole-- and the particle--spectral function are displayed
as functions of the excitation energy $\omega$ (in units of $t$) and the 
linear momentum $k$ in units of $(2\pi/N)$. The spectral functions
have been obtained here by using the $m=5$ determinants corresponding
to the five lowest LMSPHF--solutions in Fig.~\ref{fig10}. In order to obtain
continuous functions for each state a Lorentz--shape with constant
width of 0.05 $t$ has been used.}
\label{fig12}
\end{figure}

\noindent
Also for the half--filled $N=14$--lattice, the LMSPHF-- and exact
occupation numbers of the ground state can hardly be distinguished
and are not shown here. Instead, we present
in Fig.~\ref{fig12}, the  hole-- and particle--spectral functions
obtained with the $m=5$ HF--transformations resulting from the five
lowest LMSPHF--solutions out of Fig.~\ref{fig10}. In order to obtain continuous
functions of the excitation energy $\omega\,=\,\epsilon_{\tilde h}$ (or 
$\omega\,=\,\epsilon_{\tilde p}$), each state has been broadened
with a Lorentzian of constant width 0.05 $t$. As can be seen,
the particle-- and hole--strengths are symmetrically distributed
around the Fermi--energy (again $\omega_F\,=\,U/t/2$) and separated by 
a Hubbard--gap of about the same size ($U/t/2$) as obtained for the 
half--filled $N=12$--lattice.

\subsection{Large $N$=30 lattice}

Finally, we shall report LMSPHF--results for the half--filled 
$N=30$--lattice, again for interaction strength $U/t=4$. Here in the
usual Lanczos--approach $2.9\cdot 10^{15}$ states would have to be included,
which is obviously impossible. Even coupling the configurations
to good total spin and linear momentum quantum number, still for the
various $\xi$--values dimensions of the order of $9.7\cdot 10^{13}$ for
$S=0$ and even $2.2\cdot 10^{14}$ for $S=1$ would have to be treated.
In the LMSPHF--approach, however, for each $S=0$--state ``only'' functions
of 1800 real variables have to be minimized.

\begin{figure}[htb]
%\center
\includegraphics[scale=0.5,angle=270]{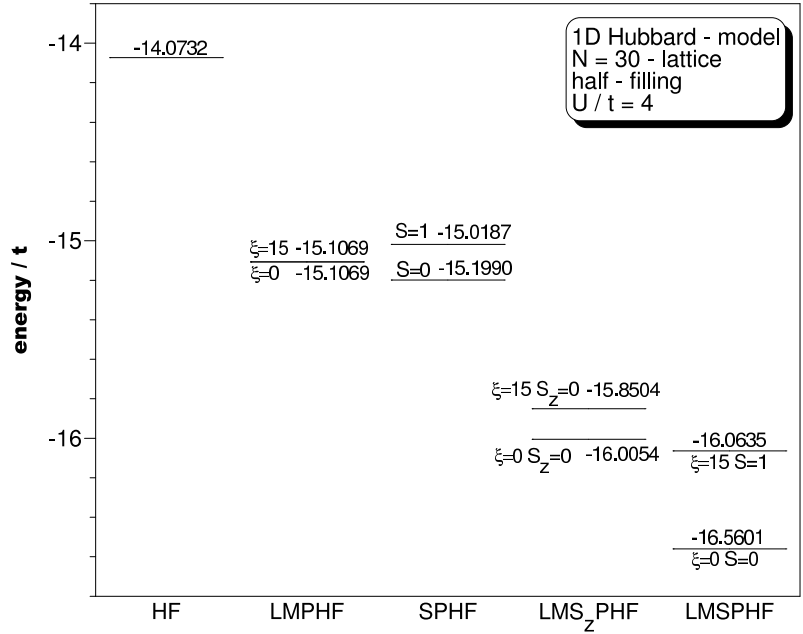}
\caption{Same as in Fig.~\ref{fig2}, but for the half--filled $N=30$--lattice.
Again $U/t=4$. Here no exact results are available to compare with. }
\label{fig13}
\end{figure}

\noindent
As in Figs.~\ref{fig2} and \ref{fig10} we present the results for the energies of the
lowest states as obtained with various variational methods for the
half--filled $N=30$--lattice in Fig.~\ref{fig13}. Obviously, there is no exact
result available to compare with. 
According to the spin coupling arguments, since $N/2$ is odd the
ground state has $S=0$ and linear momentum quantum number $\xi=0$ while
the first excited state is obtained for $S=1$ and $\xi=N/2=15$. Again
it is seen that the simultaneous restoration of linear momentum and
total spin before the variation as done in the LMSPHF--approach is
essential. All other approximations produce results, which 
do not come close to the LMSPHF--energies.

\begin{figure}[htb]
\center
\includegraphics[scale=0.5,angle=270]{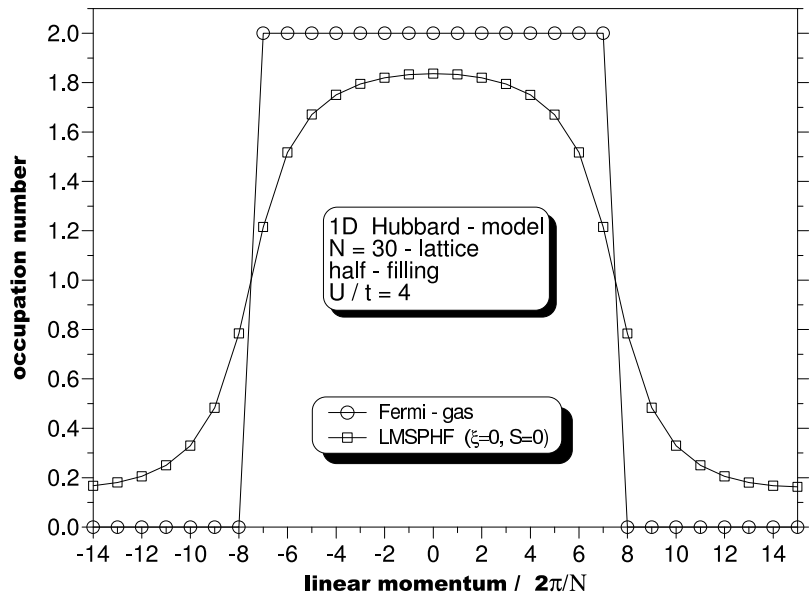}
\caption{Same as in Fig.~\ref{fig4}, but for the half--filled $N=30$--lattice.
Again $U/t=4$, and no exact results are available. }
\label{fig14}
\end{figure}

\noindent
The occupation numbers for the different basis states~(\ref{eq3}) in the
LMSPHF--ground state are presented in comparison with the values
expected for a non--interacting Fermi--gas in Fig.~\ref{fig14}. 
Like for the half--filled $N=12$ lattice 
(see Fig.~\ref{fig4}), strong effects of the correlations are observed.

\begin{figure}[htb]
\center
\includegraphics[scale=0.5]{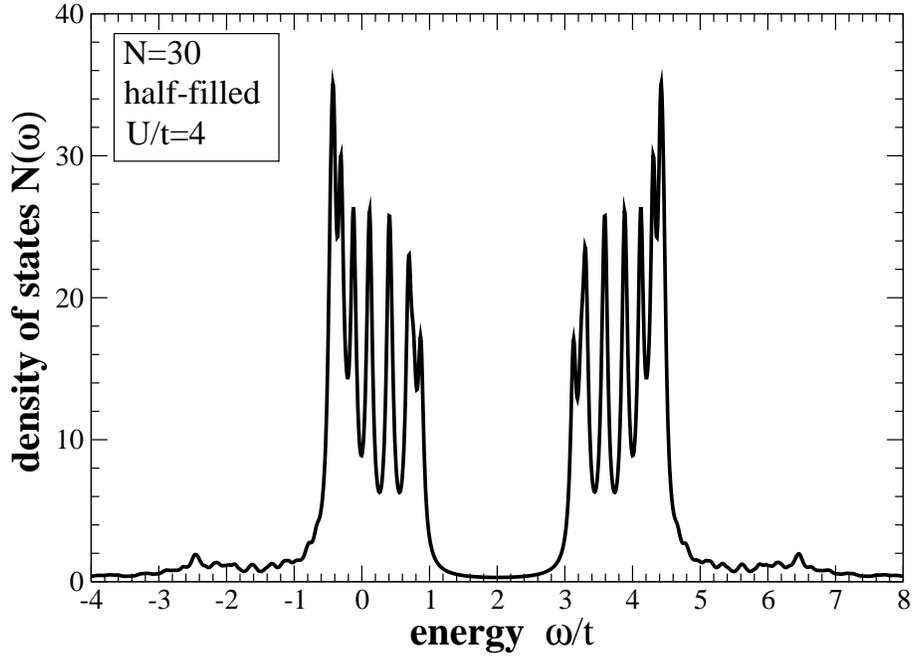}
\caption{For the half--filled $N=30$--lattice with $U/t=4$ the
total number of states $N(\omega)$ out of eq.~(\ref{eq68}) is presented as
function of the excitation energy. Here only $m=1$ determinant
was used to obtain the one--hole-- and one--particle states. Each of
these states has been artificially broadened to a Lorentz--shape
with constant width of 0.05 $t$.}
\label{fig15}
\end{figure}

\noindent
The density of states $N(\omega)$ (eq.~(\ref{eq68})) for the half--filled
$N=30$--lattice is shown in Fig.~\ref{fig15}. 
Here, we have performed a simple 1--determinant calculation ($m=1$) as
  explained in sec.~\ref{ssec23}. Again, the states were
broadened by a Lorentzian with a width of 0.05~$t$.
As expected for half--filling, particle-- and hole-- strengths are
distributed symmetrically. The Fermi--energy is again 
$\omega_F\,=\,U/t/2$ and the width of the Hubbard--gap again of about the
same size as obtained for the half--filled $N=12$-- and
$N=14$--lattices (see figs.~\ref{fig6} and \ref{fig12}).
.

\begin{figure}[htb]
\center
\includegraphics[scale=0.5]{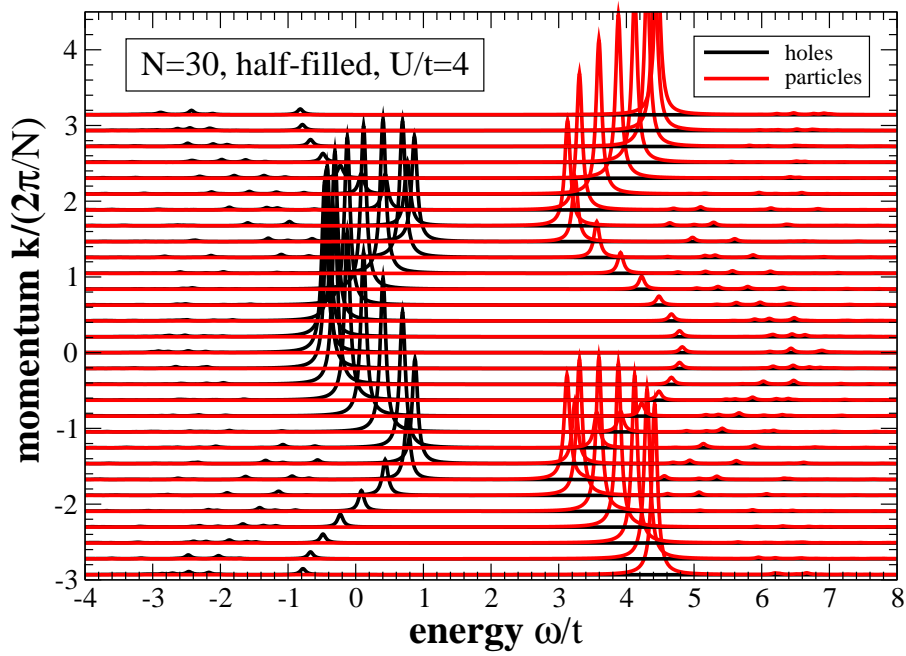}
\caption{(Color online) Same as in Fig.~\ref{fig12}, but for the half--filled
$N=30$--lattice, and using only $m=1$ determinant in the calculation of the
spectral functions.}
\label{fig16}
\end{figure}

\noindent
Finally, in Fig.~\ref{fig16}, the  particle-- and hole--spectral
functions are presented as functions of linear momentum $k$ and
 excitation energy $\omega$. Again, Lorentz--shape and a constant
width of 0.05 $t$ has been used. Obviously, the strength of each of these
spectral functions would still be redistributed, if instead of only
one, several determinants would be taken into account for the
calculation. Since anyhow, there are no ``exact'' (or experimental) data to
compare with, this generalization has not been done for the $N=30$--lattice
in the present investigation.

\section{CONCLUSIONS\label{sec4}}

We have devised a variational approach for the approximate solution
of the 1--dimensional Hubbard--model with nearest neighbor hopping and
periodic boundary conditions. For the ground state we start with a
Hartree--Fock type transformation mixing all the quantum numbers of
the single electron basis states. 

\noindent
The results may be summarized as follows~:

\noindent
For many--electron systems on small lattices, where the number of
configurations to be treated for each pair of good spin-- and linear
momentum--quantum number is smaller than the number of real variational
degrees of freedom accounted for in the variational calculation, the
LMSPHF--approach is obviously exact. However, even in the half--filled
$N=12$-- and $N=14$--lattices, where for each pair of quantum numbers
($S,\,\xi$) the number of configurations is by 2 or even 3 orders of
magnitude larger than the number of variational degrees of freedom,
the LMSPHF--approach does reproduce the exact energies of the ground and
lowest excited states with an accuracy of better than 99 percent,
the occupation numbers of the ground and lowest excited states even
better, and yields (at least if several determinants for the calculation
of the one--hole and one--particle spectra are included) even for the
spectral functions very good agreement with the exact results.
This gives some confidence, that the LMSPHF--approach can be
considered as a very good truncation scheme even for lattices which
are far too large to allow for exact diagonalisation.
Even where complete diagonalisation via the Lanczos method is
still numerically possible, the LMSPHF--approach is much faster. 

\noindent
Spin-- and linear momentum-- quantum numbers for the ground and the
lowest excited state can reliably be predicted making use of the
fact that for two electrons with opposite spin--directions in two
different momentum space basis orbits always the $S=1$--configuration
is energetically favored with respect to the $S=0$--one. Thus,
for half--filled lattices with $N/2$ being odd the ground state
has $S=0$ and linear momentum $\xi=0$, while the lowest excited
state has $S=1$ and $\xi=N/2$. For half--filled lattices with
$N/2$ even on the other hand one obtains $S=0$ and $\xi=N/2$
for the ground and $S=1$ and $\xi=0$ for the first excited state.
The same spin--correlation can be used to predict the quantum
numbers of the lowest states away from half--filling and is
also supported by the observation that for almost all linear momentum 
quantum numbers (except for that of the ground state) the lowest
state has always total spin $S=1$.

\noindent
One of the nicest features of the LMSPHF--method is that it still can be
improved rather easily. So, e.g., instead of HF--determinants generalized
Slater--determinants of the HF--Bogoliubov (HFB) type can be used
as basic building blocks. This would increase the number of real variational
degrees of freedom (for the example of an $S=0$--state) from $2N_e(2N-2N_e)$
in the LMSPHF--approximation to $2N(2N-1)$ for any number of electrons. The
price one has to pay for this, is an additional integration due to the
(then necessary) projection onto good total electron number in the
calculation of all the matrix--elements. Because of the simple form of the
interaction, however, this should not cause any serious problems. The
resulting linear momentum--, spin-- and number projected HFB--approach
would consequently account for considerably more correlations in each
single configuration than the LMSPHF--method.

\noindent
Furthermore, instead of using essentially only one symmetry--projected
configuration for each state, correlating symmetry--projected configurations
can be added and the underlying HF-- (or HFB--) transformations again be
determined by variational calculations. Since the energy gain obtained for
a particular state under consideration due to the last added configuration
is by construction always smaller than the energy gain due to the last
configuration added before, this procedure (which already has been applied
successfully to the nuclear many--body problem~\cite{sch04}) can also
give a hint on the quality of the solution even in lattices, which are
too large to allow for an exact solution.

\noindent
Last but not least, the procedure discussed in the present work can
be extended easily to the 2--dimensional Hubbard--model, which is
supposed to be of larger physical relevance than its 1--dimensional
simplification. Then, obviously, the total linear momentum becomes a
2--dimensional vector and one has to project on each of its components
separately. Thus again, with respect to the LMSPHF--approach discussed
above only one additional integration is needed.

\noindent
This leaves ample space for future investigations and we are quite
confident that the variational approaches originally devised for
the nuclear many--body problem will turn out to be rather useful
even for more complicated lattice Fermion models than the simple
1--dimensional version investigated in the present work.

\begin{acknowledgments}
We are grateful that the present study has been supported by a
grant from the Ministry of Science, Research and the Arts 
of the state Baden--W\"urttemberg (Az : 24-7532.23-19-18/1 and /2) 
to JM via the Landesforschungsschwerpunkt ``Quasiteilchen''.
\end{acknowledgments}

%\listoffigures

\newpage

\end{document}